\documentclass[a4paper,11pt]{article}
\usepackage{jheppub}

\usepackage{epsfig,amssymb,bbm,euscript,xspace,xcolor,setspace,amsmath,mathtools,empheq,amsthm,graphicx,mathrsfs,float}
\usepackage{hyperref}
\usepackage{subfig}
\usepackage{makecell,multirow} 
\usepackage{bbding}
\usepackage{color}
\usepackage{xcolor}
\usepackage{tcolorbox}
\hypersetup{
    colorlinks=true,
    linkcolor=black,
    citecolor=blue
    } 
\usepackage{comment}
\usepackage{wrapfig}
\usepackage{verbatim}
\makeatother

\preprint{MPP-2025-226, TUM-HEP-1584/25}

\title{Approximating Feynman Integrals Using Complete Monotonicity and Stieltjes Properties}

\author[a,b]{Sara Ditsch,}
\author[a]{Johannes M. Henn,}
\author[a]{Prashanth Raman}

\affiliation[a]{Max-Planck-Institut für Physik, Werner-Heisenberg-Institut, Boltzmannstr. 8, 85748 Garching, Germany}

\affiliation[b]{Technical University of Munich, School of Natural Sciences, Physics Department,  James-Franck-Straße 1, 85748 Garching, Germany}

\emailAdd{ditsch@mpp.mpg.de}
\emailAdd{henn@mpp.mpg.de}
\emailAdd{praman@mpp.mpg.de}

\abstract{
We introduce two novel numerical approaches for computing Feynman integrals based on their complete monotonicity (CM) and Stieltjes properties.
The first method uses that scalar Feynman integrals are CM, meaning that all their derivatives have a fixed sign, in the Euclidean kinematic region. This imposes strong constraints on the function space.
Simultaneously, these integrals obey systems of linear differential equations with respect to kinematic parameters.
By imposing that the solutions to these differential equations satisfy complete monotonicity across the Euclidean region, we develop an efficient and highly constraining numerical bootstrap method. 
We provide a proof of principle of the power of our approach by applying it to a class of multi-loop Feynman integrals with internal masses. 
The second method is based on a refinement of CM. We prove that Feynman integrals, within a certain range of parameters, such as dimension and propagator exponents, are not only CM but in fact Stieltjes functions. The latter can be described efficiently by Padé approximants that are known to converge in the cut complex plane. This means that these representations are valid also in analytically continued kinematics, such as physical scattering regions.
These insights allow us to obtain rational approximations to Feynman integrals from minimal information, such as a Taylor expansion about a soft limit. We demonstrate the effectiveness of this method by applying it to a 20-loop banana-type Feynman integral.
Finally, we comment on a number of extensions of these novel avenues for computing Feynman integrals.}

\begin{document}

\setcounter{tocdepth}{2}

\maketitle

%%%%%%%%%%%%%%%%%%%%%%%%%%%%
%
%
%
%
%
%%%%%%%%%%%%%%%%%%%%%%%%%%%%

\section{Introduction}

Feynman integrals are crucial components in the computation of observables in perturbative quantum field theory, in particular for scattering amplitudes relevant in collider physics and gravitational wave physics \cite{Travaglini:2022uwo,Huss:2025nlt,Buonanno:2022pgc}. In recent years, the development of new mathematical structures has significantly advanced our ability to compute and understand Feynman integrals, see for instance ref.~\cite{Abreu:2022mfk}.
However, the full analytic evaluation of multi-loop Feynman integrals remains a challenge, especially when multiple scales are involved. At the same time, there is considerable interest in numerically evaluating observables.
It is therefore interesting to ask: How can we leverage analytic insights for numerical evaluation of Feynman integrals?

In the literature, a number of promising methods exist, as we review presently. 
These include methods based on contour deformation and sector decomposition \cite{Binoth:2000ps} of Feynman integrals in Feynman parametrization. Sector decomposition helps to disentangle overlapping singularities by reorganizing the integral into a form in which divergences become more transparent and easier to handle numerically. 
See for refs.~\cite{Heinrich:2023til,Smirnov:2015mct} for computer implementations of this method. 
The authors of ref.~\cite{Borinsky:2023jdv} use tropical Monte-Carlo-based sampling methods. This method, while powerful and performant, is currently available for finite integrals only.

Another approach is to derive and to numerically evaluate differential equations satisfied by Feynman integrals.
For example, refs.~\cite{Hidding:2020ytt,Armadillo:2022ugh,Prisco:2025wqs,Mezzarobba2025Dynaverse} rely on knowing the value of the integrals at some reference point (e.g. as determined by physical consistency conditions, or via other means, such as numerical evaluation via the above methods), and then analytically continue that value to other kinematic values by matching the domains of convergence of series expansions about different kinematic points. For this to work well, it is desirable to have the differential equation in a form as simple as possible (cf. the review \cite{Henn:2014qga}), which requires additional steps. 
On the other hand, ref.~\cite{Liu:2022chg} introduces an auxiliary parameter, with the help of which it becomes easy to determine boundary values of an integral, which are then analytically continued to the integral without the auxiliary parameter. While very powerful and universally applicable, introducing the additional scale of the auxiliary parameter sometimes constitutes a bottleneck.  

While the methods discussed above can be very useful, they
also come with drawbacks, which limit their reach. The sector decomposition approach requires a time- and memory-consuming preparation phase. Moreover, its applicability is often restricted to integrals of moderate number of loops and legs.
While obtaining differential equations for Feynman integrals is in principle a standard task that can be automated, this comes with important computational bottlenecks. 
This motivates us to investigate alternative approaches.

In this paper, we introduce two new ideas for constraining and determining Feynman integrals numerically, which are based on powerful mathematical properties that so far have not received much attention in the multi-loop research community. 

The first approach we consider also takes differential equations as input, but does not require any specific form of them, which makes our approach easy to use and without complicated prerequisites. 
More specifically, we propose a new method for bootstrapping Feynman integrals, by combining their differential equations with the constraint of complete monotonicity. Complete monotonicity is a powerful positivity condition on (all) derivatives of a function. In ref.~\cite{Henn:2024qwe} two of the present authors showed that many relevant quantities in quantum field theory are completely monotone. See refs.~\cite{Ma:2025pvq,Mazzucchelli:2025gyg} for further recent research in this direction.  

In particular, scalar Feynman integrals exhibit complete monotonicity in their Euclidean region with respect to kinematic invariants \cite{Henn:2024qwe}. 
When combined with differential equations for a set of integrals, this property imposes powerful constraints on the space of possible solutions. We show that this interplay between differential equations and complete monotonicity 
can uniquely determine the integral. This extends previous work on numerically bootstrapping Feynman integrals \cite{Zeng:2023jek}, which employed positivity properties of functions, but did not consider derivatives.

The second approach uses powerful mathematical properties that hold for Stieltjes functions, which are a subset of CM functions.
We prove that Feynman integrals are Stieltjes, when their parameters (i.e. propagator powers, dimension) lie within a certain range.

Stieltjes functions enjoy remarkable analyticity properties that imply that they can be extremely well described by Padé approximants~\cite{Basdevant1972Pade,BGM}. In particular, strong bounds and convergence theorems are known for the Padé approximants. Leveraging this knowledge, we show how one can obtain reliable numerical approximations over a wide range of kinematic space, including complex-valued, analytically continued results in physical regions.
This allows for a fast and high-precision numerical determination of integrals across a wide range of kinematics by exploiting minimal analytic assumptions to obtain tight numerical constraints.

In this paper, we provide the theoretical background to these ideas in the context of Feynman integrals, as well as a proof-of-principle application to a class of multi-loop integrals that effectively depend on one variable. For technical convenience, we chose a class of massive multi-loop propagator integrals. 

It is important to emphasize that the Stieltjes properties discussed above are also useful without knowledge of a differential equation. We show how one can obtain an excellent approximation to Stieltjes Feynman integrals from the knowledge of their Taylor expansion around one kinematic value. As a case in point, we demonstrate this for a 20-loop banana-type integral.

The outline of the paper is as follows. In Section~\ref{sec:2}, we introduce the main cast of characters to the bootstrap method, namely completely monotone functions and differential equations for Feynman integrals.
We present the pedagogical example of the massive bubble integral to illustrate the method and to highlight its key features. We then proceed to multi-loop applications.
In Section \ref{sec:beyondiff}, we introduce Stieltjes functions and explain their relation to Padé approximations. We provide a proof that a class of Feynman integrals is Stieltjes (see Section \ref{sec:FeynStielt}), and give an example relevant to one-loop five-particle scattering (in Section \ref{sec:subsecPadeexample}).
We then illustrate the power of the good convergence properties of Padé approximants to Stieltjes functions. This is done in one case  in combination with the CM bootstrap (see Section \ref{sec:PadepowerCM}), and in a second case independently of differential equations. As a highlight, we determine numerically, across the complex kinematic plane, a 20-loop banana-type integral (see Section \ref{sec:twenty-loop-banana}).
Finally, in Section~\ref{sec:conclusion}, we summarize our results and provide a detailed discussion of promising future directions.

%%%%%%%%%%%%%%%%%%%%%%%%%%%%
%
%
%
%
%
%%%%%%%%%%%%%%%%%%%%%%%%%%%%

\section{The complete monotonicity bootstrap from differential equations}
\label{sec:2}

\subsection{Complete monotonicity and Feynman integrals}

A real-valued function $f: R \to \mathbb{R}$ is said to be completely monotonic (CM) in the region $R$ if it satisfies (cf. ref.~\cite{widder1941laplace})
\begin{equation}    \label{eq:CM}
    (-1)^n \frac{d^n}{dx^n} f(x) \geq 0\,, \quad \forall \, n \in \mathbb{N}_0, \; \forall \, x \in R \,.
\end{equation}
This property implies in particular that $f(x)$ is positive, decreasing, and convex. 
It is known that CM functions have characterizations via integral representations with non-negative kernels. For example, for $R= {\mathbb R}{_+}$, one has
\begin{align}\label{eq:Laplacerepr}
f(x) = \int_0^\infty e^{- x t } \mu(t) dt\,, \quad\quad {\rm with}\quad \mu(t) \ge 0.
\end{align}
There is a straightforward connection to Feynman integrals that was uncovered in ref.~\cite{Henn:2024qwe}.
In that reference, two of the present authors showed that scalar Feynman integrals are CM functions of appropriate linear combinations of kinematic variables, within their Euclidean region. By Euclidean region we mean the region in the space of kinematic variables $\{x_i\}variables = \{-s_{T,R}, m_i^2\}$ where the second Symanzik graph polynomial $F$, as defined in Appendix~\ref{sec:Feynman}, is non-negative for all non-negative values of the Feynman parameters. Note that we do not require this kinematic configuration to be realized via four-dimensional real-valued momenta.

It is well known that Feynman integrals satisfy linear differential equations, called Picard-Fuchs equations, in their kinematic invariants, see e.g. ref.~\cite{Muller-Stach:2012tgj}. Alternatively, they can be viewed as first-order matrix systems of equations, as common in the high-energy physics literature on Feynman integrals, see e.g. ref.~\cite{Henn:2014qga}. Let us adopt the latter point of view.
In the case of a single kinematic variable $x$, these equations take the following form,
\begin{equation} \label{eq:DE}
\frac{d}{d x} {\bf f}(x) = A(x) {\bf f}(x) \,,
\end{equation}
where ${\bf f}(x)=\{f_1(x),\dots,f_k(x)\}$ is a vector of basis integrals and $A(x)$ is a matrix.
For certain natural basis choices, such as the ones we will make in this paper, the entries of $A$ are rational functions of $x$. (See e.g. ref.~\cite{Henn:2014qga} for more details.)
In the following, we will choose basis integrals ${\bf f}(x)$ that are CM. 
This is expected to be always possible, as we will illustrate in section \ref{sec:multiloop}.
In fact, as we explain in that section, we can find more CM integrals than strictly needed for our present purposes.

\subsection{A linear program for bootstrapping Feynman integrals}

We can rewrite the CM property, 
eq.~(\ref{eq:CM}), in the context of the differential equation, eq. (\ref{eq:DE}), as follows,
\begin{align}
\label{eq:constraint}
    Q_n(x){\bf f}(x)\geq 0\,,\quad \forall n{\in} \mathbb{N}_0 \,,
\end{align}
where the first two cases are given by
\begin{align}
    Q_0=\mathbf{1}\,, \quad Q_1(x)=-A \,, 
\end{align}
and higher derivative matrices are obtained recursively via
\begin{align} 
    Q_n(x)=- \partial_x Q_{n-1}(x)+ Q_{n-1}(x)Q_1(x) \,.
    \label{eq:recursiondericative}
\end{align}
We view the set of inequalities given in eq. (\ref{eq:constraint}) as constraints on the possible function values at some argument $x$.
Note that these equations are  invariant under rescaling ${\bf f}(x)$ by any $\lambda >0$.
Therefore, one cannot fix the overall normalization from these equations alone.
This is a fact that is well known from differential equations for Feynman integrals.
However, just as there, one can take certain known elementary basis integrals to fix the overall normalization.
We comment on how this can be done in practice in section \ref{sec:multiloop}.

Let us now describe the bootstrap procedure in more detail.
The input, apart from $Q$, is a number $n_0$ and a point $x_0$.
We then compute all $Q_n$ with $n \le n_0$, and evaluate them at $x=x_0$. 
Then eq. (\ref{eq:constraint}) provides numerical constraints on ${\bf f}(x_0)$. 
To obtain the feasible region for the numerical value of the basis integrals ${\bf f}(x_0)$, we set up a {\it linear program}.
In other words, after normalizing one of the components of ${\bf f}(x)$, say without loss of generality $f_1(x
)=1$, we maximize and minimize the other components $f_j(x)$ with $j\neq 1$, subject to the constraints of eq. \eqref{eq:constraint}. 
As we will see in section \ref{sec:pedagogocalexample} with the help of a pedagogical example,
depending on the value $x_0$, this gives us constraints of the function value ${\bf f}(x_0)$. In the best possible case, 
we obtain tight, two-sided, bounds that converge as $n_0$ is increased. 

It is worth emphasizing that for our purpose, it is not required to find particularly simple, or e.g. canonical \cite{Henn:2013pwa}, representations of the differential equations matrix. This means that there are few prerequisites, other than availability of some form of the differential equations matrix. Also, note that if scalar CM integrals are chosen as a basis, this means that the entries of the differential equations matrix are rational, which is advantageous in view of computer algebra.

%%%%%%%%%%%%%%%%%%%%%%%%%%%%
%
%
%
%
%
%%%%%%%%%%%%%%%%%%%%%%%%%%%%

\begin{figure}[t]
\centering
\includegraphics[width=0.4\linewidth]{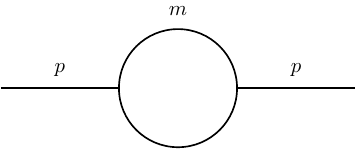}
\caption{The one-loop massive bubble Feynman integral.}
\label{fig:bubble}
\end{figure}

\subsection{Pedagogical example}\label{sec:pedagogocalexample}

In this section, we apply the theoretical framework developed in the previous section to numerically determining Feynman integrals. In order to do so, we combine the new insights about the relationship of Feynman integrals to completely monotone and Stieltjes functions, with state-of-the-art differential equations method. To make this part accessible and easy to follow, we illustrate all ideas with the help of a pedagogical example.

\subsubsection{Differential equations}
\label{sec:de}

We consider the two-dimensional bubble integral with an internal mass, cf. Fig.~\ref{fig:bubble},
\begin{align}
f(x) = \int \frac{d^{2}k}{i \pi} 
\frac{1}{
(-k^2+m^2) 
[-(k+p)^2+m^2] 
  } \,.
\end{align}Without loss of generality, we set $m=1$, and denote $x =-p^2$. 
In the following, we will consider the CM property w.r.t. $x$. 
In principle, the CM property in $m$, which also holds, could give additional restrictions, but we do not consider them here, for simplicity of presentation.

The integral $f(x)$ can be evaluated in the Feynman parametrization, cf. chapter 4 of ref.~\cite{Badger:2023eqz}. The result is
\begin{align}\label{bubbleans}
    f(x) =   \tfrac{2}{x \sqrt{
   (1+ 4/x)}} \log \left( \tfrac{1 + \sqrt{1 + 4/x}}{-1 + \sqrt{1 + 4/x}} \right)\,.
\end{align}
We will use this for numerical reference values.
The Euclidean region of the bubble integral corresponds to $x>-4$, see Appendix~\ref{sec:Feynman} for details.
Indeed, $f(x)$ is completely monotone in $x$ in that region.

\begin{figure}[t]
\centering 
\includegraphics[width=0.7\textwidth]{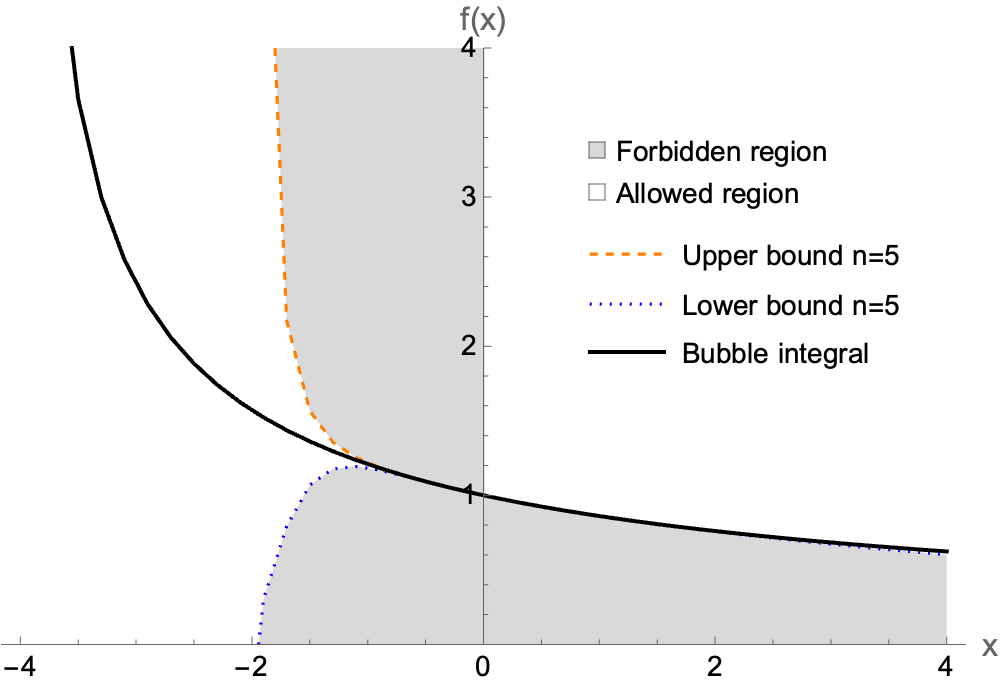}
\caption{Constraints obtained from the interplay of differential equations and complete monotonicity for the massive bubble integral.
The dashed yellow and dashed blue lines correspond to upper and lower bounds, respectively. 
The white regions and gray regions indicate allowed and forbidden regions, respectively.
The solid black line corresponds to the exact function value.}
\label{fig:bubblebootstrap}
\end{figure}

The bubble family has two basis integrals, which we denote as 
\begin{align}
{\bf f} = \{  1, f\} \,.
\end{align}
The first integral is the tadpole integral, which depends on the mass $m$ only, and not on the parameter $x$. We therefore can fix its value to a constant. 
The differential equation for these basis integrals in $D=2-2~ \epsilon$ reads
\begin{align}\label{DEMatrices}
\partial_{x} {\bf f} = A_x \, {\bf f} \,, \quad  {\rm with} \quad
  A_{x} = -\begin{pmatrix}
0 & 0 \\
-\frac{2 }{(4 +x ) x} & \tfrac{2  +x+ ~\epsilon~ x}{(4 +x) x}  
\end{pmatrix}\,.
\end{align}

\subsubsection{Numerical bootstrap}
\label{sec:CMbootstrap}

We apply the linear program outlined in the previous section, to obtain constraints on the integral value in the Euclidean region. The results for $n_0=5$ are shown in Fig.~\ref{fig:bubblebootstrap}. We summarize them as follows:
\begin{itemize}
\item $x \in (-4,-2)$:
We get only a lower bound (dotted, blue) that does only very loosely constrain the value of the function.
\item $x \in (-2,0)$:
We get both a lower and an upper bound (dashed, yellow), which tightly constrain the value of the function.
\item $x \in (0,\infty)$: We get only a lower bound. However, very interestingly, that bound is very close to the value of the function (solid, back).
\end{itemize}

This change in behavior between the regions is related to the fact that some entries of the differential equation matrix $A_x$  change sign or become singular at the boundary of the regions.

In the region with two-sided bounds, the value can be estimated by the mean of the upper and lower bounds, with the error given by the distance to the bound. 
In Table \ref{tab:bubbleconvergence}, we show the convergence in the region with a two-sided bound.
In general, we find that the value converges quickly against the value of the function, and gives a good approximation already when including just a few derivatives, e.g. for $n_0=5$. The closer we get to the upper bound at $x=0$, the better the convergence, whereas the values converges only slowly closer to the lower bound at $x=-2$.

\begin{table}[t]
    \centering
    \begin{tabular}{c|ccc}
         $x_0$ &-1.55&-0.66&-0.32  \\
    \hline
             $n_0=5$&1&1.12& 1.05 \\
         
         $n_0=15$&1.3&1.126928042&1.05699867495761\\
         $n_0=30$&1.3792&1.126928042783530&1.056998674957614\\
        $f(x_0)$ &1.37921814937616 &1.126928042783530  &1.056998674957614 \\
           \end{tabular}
    \caption{ 
    We provide the numerical value for the integral up to the number of digits that can be obtained from the CM bootstrap including $n_0$ derivatives for points in the region with two-sided bounds. If the precision is higher than $15$ digits, we cut at 15 digits.} 
    \label{tab:bubbleconvergence}
\end{table}

%%%%%%%%%%%%%%%%%%%%%%%%%%%%
%
%
%
%
%
%%%%%%%%%%%%%%%%%%%%%%%%%%%%

\subsection{Multi-loop applications}
\label{sec:multiloop}

In this section, we discuss applications of the method to the equal-mass `banana' integrals, see Fig.~\ref{fig:Stieltjes_pdf}(a), up to four loops. These integrals are relevant to phenomenology as subtopologies for processes with internal massive propagators. They provide an optimal playground for a numerical bootstrap, as they are analytically complicated with an easy to study differential equation and convergence behaviour. They are known to be associated with elliptic curves and other Calabi-Yau manifolds \cite{Bloch:2014qca}, which makes their function space and analytic computation more involved. The CM bootstrap does not depend on the analytic complexity of the function space, and efficiently provides numerical values also for these more complex functions. The differential equation for these integrals is known at all loop orders and was solved up to six loops, see refs.~\cite{Lairez:2022zkj,Pogel:2022vat}. 
We demonstrate the efficient numerical evaluation of the equal-mass `banana' integral families up to four loops. We discuss the runtime for different precision goals and compare the performance to other commonly used methods for numerical evaluation.

\subsubsection{Choice of a suitable CM basis}
 We work in $D=2$ spacetime dimensions and use higher powers of propagators for sectors with several basis integrals. Finite scalar Feynman integrals are CM in the Euclidean region, as shown in ref.~\cite{Henn:2024qwe}. Therefore we choose basis integrals that are ultraviolet (UV) and infrared (IR) finite: The UV behaviour of a scalar $L$-loop integral  in $D$ spacetime dimensions with propagator powers $\{\nu_i\}$ is governed by the superficial degree of divergence 
 \begin{equation}
     \omega = L\cdot D - 2\sum_i \nu_i\,,
 \end{equation}together with the corresponding degrees for all of its subgraphs. The integral is UV finite if the superficial degrees of divergence of the graph itself as well as of all the subgraphs are negative. This implies that the one-loop integrals subgraphs as well as the full higher-loop `banana' integrals are already UV convergent. The only potential UV divergences arise from single-propagator one-loop, i.e. $L=1$, subgraphs, so-called tadpoles, since their superficial degree of divergence is  $\omega_\text{tad}=D-2\cdot 1=0$ in $D=2$ dimensions and therefore generate a logarithmic UV divergence. This occurs, for example, in the subsector $I_{1,1,0}$ of the two-loop `banana' integral, which factorizes into a product of two tadpoles.  By choosing a power $2$ or higher for all tadpole propagators, i.e. $\nu_i\geq 2$, the superficial degree becomes negative, $\omega'_\text{tad} \geq 2 - 2\cdot 2 = -2$, and all potential UV subdivergences are removed. Moreover, the integrals are trivially IR finite, as all propagators are massive, meaning that they take the form $(k^2+m^2)^{-1}$, which stays finite for $k\to 0$.

Our choice of basis integrals as well as the definition of the integral families are provided in ancillary files.

\subsubsection{CM bootstrap and results}

\begin{table}
\centering
\renewcommand{\arraystretch}{1.3}
\setlength{\tabcolsep}{10pt}
\begin{tabular}{c|ccc} 
   Family   &   \makecell{Number of \\ Basis Integrals} & \makecell{Euclidean \\ Region}  & \makecell{Two-Sided \\ Bounds}\\ 
\hline
Two-loop `banana' integral 
&   3 & $(-9,\infty)$  &$(-3,-1)$ \\
Three-loop `banana' integral &  4 & $(-16,\infty)$  & $( \alpha_1,-4)$ \\
Four-loop `banana' integral&  5  & $(-25,~\infty)$   & $(\alpha_2,-9)$\\
\end{tabular}
\caption{Results of the region analysis in the context of the CM bootstrap for the equal mass `banana' integrals up to four loops. The lower bound of the two-sided region at three and four loops is given by $\alpha_1 \approx -8.51$, which is one of the roots of the equation $ 256 +264x+36x^2+x^3 =0$ and $\alpha_2 \approx -12.17$, one of the roots of $ 7650 + 8645x + 1616x^2 + 103x^3 + 2x^4=0$.}
\label{resultsBanana}
\end{table}

Having determined a set of suitable basis integrals, we can now perform the CM bootstrap as introduced in Section \ref{sec:CMbootstrap}. For each of the families, we first compute the differential equation as explained in Section \ref{sec:de}. We use \texttt{FiniteFlow}~\cite{Peraro:2019svx} in combination with \texttt{Kira}~\cite{Maierhofer:2017gsa} to obtain and solve the relevant IBP relations. The resulting differential equations for each family are provided in ancillary files.

Next, we determine the zeros and poles of the differential equation matrix in the Euclidean region, identifying regions within which the quality of the bound is not expected to change. Within each region, we select a random point using the \texttt{Mathematica} function \texttt{FindInstance} and run the linear program introduced in Section \ref{sec:CMbootstrap} to obtain bounds on the integral value at this point. This allows us to identify the regions with two-sided bounds for each family.

We summarize our results for the CM bootstrap in Table \ref{resultsBanana}. 
For each integral family, we indicate the Euclidean region, namely $(-(L+1)^2,\infty)$ for the $L$-loop `banana' integral, 
where the CM property holds, 
as well as the intervals with two-sided bounds. 
We find that for the `banana' integrals, the situation is very similar to the bubble example. There is a region with two-sided bounds close to the lower end of the Euclidean region, while in the rest of the Euclidean region there is a lower bound only.

\subsubsection{Performance analysis}
\begin{figure}
    \centering
    \includegraphics[width=0.9\linewidth]{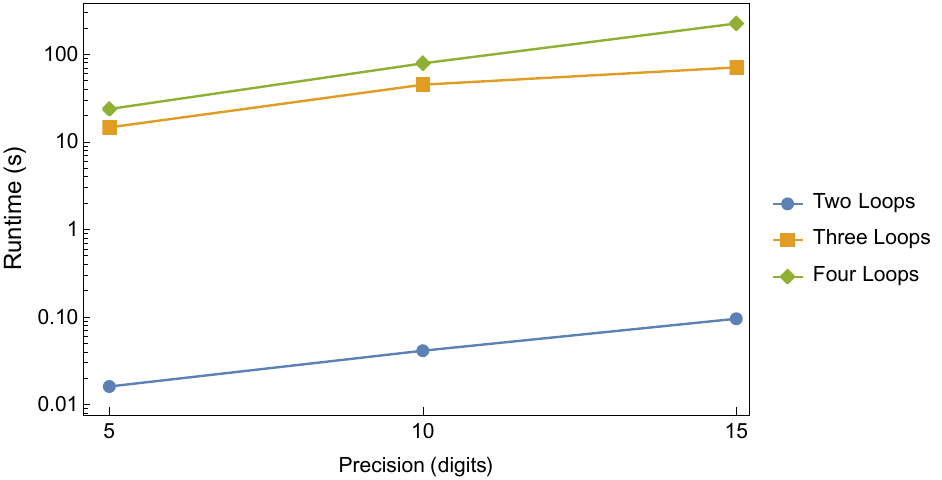}
    \caption{Average runtime of the CM bootstrap for the `banana' integrals at different loop orders with different precision goals.}
    \label{fig:PerformancePrecision}
\end{figure}
We first analyze how the runtime scales with the number of digits obtained. For that purpose, we randomly generate ten points using the \texttt{Mathematica} function \texttt{RandomReal} within the region with a two-sided bound for each family and compute the values of the integrals for different precision goals. We show the average runtime for different precisions in Fig.~\ref{fig:PerformancePrecision}.\footnote{Numerical results were obtained on a dual-socket AMD EPYC 7763 node (128 physical cores, 256 threads, 1.65 TB RAM).}

We find that the runtime increases with higher precision as well as with a higher number of loops. Both of these are expected, as for higher precision, we need to include more derivatives and at more loops the number of basis integrals increases, both of which leads to more inequalities that have to be solved, which in turn results in a longer runtime. 

We also find that, just as for the bubble integral (cf. Table \ref{tab:bubbleconvergence}), the convergence of the bounds and therefore the runtime depend heavily on the phase space point. We find that the points closer to the lower end of the region with two-sided bounds require more derivatives than those close to the upper end. For example, to obtain ten-digit precision for the three-loop `banana' integral, the point closest to the lower end ($x=-7.78$) required 429 derivatives, leading to a runtime of 158.8\,s, while the point closest to the upper end ($x=-4.02$) required just 27 derivatives which resulted in a runtime of 0.200\,s. This is a difference of three orders of magnitude, that is caused only by the choice of phase space point.

We also compare the performance of our method against a commonly used program for the numerical evaluation of Feynman integrals. For convenience, we chose \texttt{AMFlow} \cite{Liu:2022chg}. For the comparison, we compute the integrals at ten digit precision. This is a reasonable precision goal for various practical applications of numerical Feynman-integral evaluation, where phenomenological uncertainties (parton distribution functions, parameters, scale variation) typically dominate over the numerical integration error. 

We evaluate the basis integrals with \texttt{AMFlow} on the same ten randomly selected phase space points also used in the  analysis above. In Table \ref{PerformanceComparison}, we provide the average evaluation time per point with \texttt{AMFlow} compared against the CM bootstrap.\footnote{Numerical results were obtained on the same machine as for the previous analysis (cf. Footnote 1). \texttt{AMFlow} was run using 50 OpenMP threads.} We find that the evaluation time with \texttt{AMFlow} is much less point dependent than of our method. For example, the evaluation times for the three-loop `banana' vary between 22.19\,s and 28.29\,s, which is a much smaller variation as the one of the CM bootstrap, which was three orders of magnitude. 
Depending on the point, the CM bootstrap is much faster or much slower than \texttt{AMFlow}. One important difference between the  CM bootstrap and \texttt{AMFlow} is that in \texttt{AMFlow} the whole process has to be repeated for each phase space point and no intermediate steps are saved, while for the CM bootstrap we compute the differential equation and its derivatives just once analytically. While this is a time-consuming step, the fact that it has to be performed just once leads to a big speed-up when evaluating several phase space points.

\begin{table}[t]
    \centering
    \renewcommand{\arraystretch}{1.3}
    \begin{tabular}{c|ccc}
        Family & CM Bootstrap & \texttt{AMFlow} \\
         \hline
         Two-loop `banana' integral&0.041&14.35\\
         Three-loop `banana' integral&44.76&23.42\\
         Four-loop `banana' integral&78.43&126.2\\
    \end{tabular}
    \caption{Comparison of the performance of the CM bootstrap against \texttt{AMFlow}. The average runtime for the evaluation of one phase space point is shown in seconds.}
    \label{PerformanceComparison}
\end{table}

We conclude that the performance heavily depends on the position of the point within the region with two-sided bounds. For certain points close to the upper bound of the region with two-sided bounds, the CM bootstrap is very efficient in providing a numerical value for the integrals also at high precision and even outperforms established programs like \texttt{AMFlow} by orders of magnitude. 
This means that our method can be a useful tool for obtaining boundary values of integrals, for example.
Moreover, in section~\ref{sec:beyondiff}, we show that a refinement of the CM property to Stieltjes functions allows for efficient analytic continuation to generic kinematic values.

%%%%%%%%%%%%%%%%%%%%%%%%%%%%
%
%
%
%
%
%%%%%%%%%%%%%%%%%%%%%%%%%%%%
\section{Stieltjes functions and Feynman integrals}
\label{sec:beyondiff}

\subsection{Stieltjes functions}
\label{sec:stieltjes} 
By definition, a Stieltjes function $f(z)$ admits the following integral representation,
\begin{equation} \label{eq:Stieltjesdefinition}
  f(z) = \int_{0}^{\tfrac{1}{R}} \frac{\rho(u)}{1 + u z}\, du \,, 
  \quad 
 {\rm{ with}} \quad \rho(u) \ge 0 \; \forall \; u \in (0, 1/R) \,.
\end{equation}
It follows from eq. \eqref{eq:Stieltjesdefinition} that Stieltjes functions form a subset of CM functions, since their derivatives satisfy 
\begin{align}\label{eq:muforStieltjes}
\left( - \partial_z \right)^{n} f(z)
  = \int_{0}^{1/R} du\, \frac{u^{n}\, n!\, \rho(u)}{(1 + u z)^{n+1}}
  \ge 0 \,.
\end{align}
To prove that a function is Stieltjes it is sufficient to show the following properties (cf. Section 8.6 of  ref.~\cite{Bender}):
\begin{enumerate}
    \item \textit{Analyticity.}  
    $f(z)$ is analytic  in the cut complex plane.
    \item \textit{Asymptotic behavior.}  
   $f(z)$ approaches a non-negative real constant $C$ at large $z$, i.e.,   
    \begin{equation}\label{c}
        f(z) \rightarrow C \quad \text{as} \quad |z| \rightarrow \infty \,.
    \end{equation}
    \item \textit{Herglotz property.}  
    $-f(z)$ is a Herglotz function, i.e.
    \begin{align}\label{eq:herglotz}
        \Im f(z)\, \Im z < 0 \qquad \forall\, z \notin \mathbb{R} \,,
    \end{align}
    where $\Im$ denotes the imaginary part.
\end{enumerate}
Let us explain how these conditions imply the Stieltjes representation eq. (\ref{eq:Stieltjesdefinition}), following the arguments of ref.~\cite{Bender}. 
Since $f(z)$ is analytic in the cut plane $\mathbb{C}\setminus(-\infty,-R]$, we can apply Cauchy's theorem to
\begin{align}
\frac{f(w)-C}{w} \,,
\end{align}
where $C$ is the constant defined in eq.~\eqref{c}. The subtraction ensures that the integrand falls off as $|w|^{-2}$ for $|w|\to\infty$, so that the contribution from the large circular arc of the keyhole contour vanishes. Deforming the contour to wrap around the branch cut $(-\infty,-R]$ then gives
\begin{equation}
f(z)-C=\frac{1}{2\pi i}\int_{-\infty}^{-R}\frac{\text{Disc}\,f(x)}{x-z}\,dx \,.
\end{equation}
Using the Herglotz property~\eqref{eq:herglotz}, the discontinuity across the cut has a definite sign,
\begin{align}
\text{Disc}\,f(x)=2i\,\Im f(x+i0),\qquad x<-R,
\end{align}
with $\Im f(x+i0)\le0$. Defining the positive measure
\begin{align}
\rho(u)=-\frac{1}{\pi u^2}\Im f\!\left(-\frac{1}{u}+i0\right),\qquad u\in(0,1/R),
\end{align}
and changing variables $x=-1/u$, we obtain
\begin{align}
f(z)=C+\int_0^{1/R}\frac{\rho(u)}{1+uz}\,du \,.
\end{align}
The constant $C>0$ can be absorbed into the measure. This yields eq.~\eqref{eq:Stieltjesdefinition}.

We comment that an alternative characterization of Stieltjes functions due to Krein is given in ref.~\cite{Krein1977TheMM}.

Stieltjes functions possess useful analytic properties. In particular, they can be accurately modeled by Pad\'e approximants, as will be described the next subsection. Certain related convergence proofs rely on an 
 additional assumption about the existence of moments, that in some formulations  is included in the definition of a Stieltjes function (see refs.~\cite{Bender,BGM}). 
This leads to the following additional requirement.

\begin{enumerate}
\setcounter{enumi}{3}
\item \textit{Asymptotic series expansion.}  $f(z)$ admits an asymptotic expansion in the cut plane.
For example, for $z\to 0$ one has
\begin{align} \label{asyexp} f(z) \sim \sum_{n=0}^{\infty} (-1)^{n} a_{n}\, z^{n} \,, \end{align}
with positive coefficients $a_n>0$. These coefficients correspond to the moments of the measure,
\begin{equation}
a_n = \int_0^{1/R} u^n \rho(u)\,du .
\end{equation}
\end{enumerate}
The reason that we discuss this point separetely is the following. For $R>0$ the range of integration is finite and all moments are guaranteed to exist. However, when $R=0$ the integral extends to infinity and the moments need not exist. In this case the expansion around $z=0$ may involve logarithmic terms.
 In practice, we can avoid potential issues with moments by expanding around any point $z_0>0$, since the shifted function
\begin{equation}
g(z)=f(z+z_0)
\end{equation}
is Stieltjes. 

Let us illustrate this point with an example. The function
\begin{equation}
f(z) = \frac{\log z}{z-1}
=
\int_0^\infty \frac{du}{(1+zu)(1+u)} \,,
\end{equation}
does not admit finite moments at $z=0$. However,
\begin{equation}
g(z) = f(1+z) = \frac{\log(z+1)}{z}
=
\int_0^1 \frac{du}{1+zu} 
\end{equation}
does. In applications, we therefore expand about a point $z=z_0>0$, where a genuine power series representation is guaranteed.

\subsection{Padé approximation and convergence theorems}
\label{sec:subsecPadeconvergence}
The Padé approximation provides a simple and effective alternative to polynomial approximations for analytic functions. Given the Taylor series of a function $f(z)$ around a point $x_0$, the $[N, M]_{x_0}$ Padé approximant, denoted by $P^{N}_{M}(z;x_0)$, is defined as the ratio of two polynomials of degrees $N$ and $M$. This rational function is constructed so that its Taylor expansion of $f(z)$ around $x_0$ agrees with the original series up to terms of order $N + M + 1$.

Padé approximations are often employed because they have many desirable features, cf. refs. \cite{BGM,Basdevant1972Pade}. They usually have better convergence properties compared to Taylor approximations, and they allow one to analytically continue functions to obtain values outside the radius of convergence of the original series,
(cf. refs. \cite{Davies:2019nhm,Czakon:2020vql} and references therein for sample uses of Padé approximations in the context of Feynman integrals.).

Specifically, for Stieltjes functions, we benefit from powerful theorems, as we review presently.

\begin{figure}[t]
    \centering
    \includegraphics[width=0.4\linewidth]{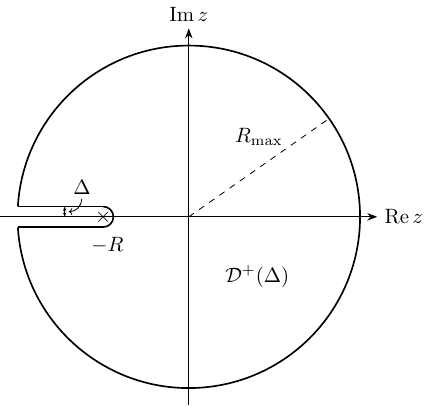}
    \caption{The domain $\mathcal{D}^+(\Delta)$ where Padé approximations are well-behaved. The convergence properties are discussed in detail in the main text.
    $R_{\text{max}}>0$ can take any value. 
    $\Delta>0$ quantifies the distance from the cut on the negative real axis.
    Figure adapted from ref.~\cite{BGM}.}
    \label{fig:pade}
\end{figure}

To set the notation, we consider functions that are analytic in complex plane apart from a branch cut along the negative real axis starting from $-R$, cf. Fig.~\ref{fig:pade}.
Padé approximations to Stieltjes functions have the following nice properties:
\begin{itemize}
\item[(a)] {\it Convergence on the real axis} (cf. Section 8.6 of ref.~\cite{Bender}.) For real arguments $x \in \mathbb{R}\backslash(-\infty,-R]$ and $x \ge x_0$, the following statements hold:
   \begin{itemize}
    \item The Padé sequence $ \{P^{N}_N (x;x_0)\}$ decreases monotonically as $N$ increases.
   \item The Padé sequence $ \{P^{N-1}_{N}(x;x_0) \}$ increases monotonically as $N$ increases.
   \item The Padé sequences  $ \{P^{N}_N (x;x_0)\}$ and $ \{P^{N-1}_{N}(x;x_0) \}$ provide an upper/lower bound to the function, respectively, 
    \begin{equation}\label{realx}
       P^{N-1}_{N}(x;x_0) \le f(x) \le  P^{N}_{N}(x;x_0), \qquad x \ge x_0 \,.
    \end{equation} 
   \end{itemize}
    \item[(b)] {\it Poles and residues} (cf. Theorem 5.2.1 in ref.~\cite{BGM}.) The poles of $  P^{N+J}_{N}(x;x_0)$ for $J\ge -1$ are all simple, lie on $(-\infty, -R)$ and have positive residues. In particular, one can write

\begin{equation}
 P^{N-1}_{N}(x;x_0) =\sum\limits_{i=1}^{N} \frac{\beta_i}{1+\gamma_i z}, \qquad\beta_i,\gamma_i \ge 0 .
\end{equation}
From this, one can see that the branch cut is approximated by the poles and zeros of the Padé approximation. 
  \item[(c)] {\it Convergence in the cut plane} (cf. Exercises 8.55, 8.59, 8.60 in ref.~\cite{Bender}.) Padé approximations have very good convergence away from the cut. In the cut plane $\mathbb{C}\backslash(-\infty,-R]$, both sequences $\{P^{N}_N (z;x_0)\}$ and $\{P^{N-1}_{N} (z;x_0)\}$ converge to the Stieltjes function $f(z)$, which has the expansion given in eq.~\eqref{asyexp}, provided that the coefficients do not grow too quickly, more precisely, if $|a_n| = O( (2n)!~ C^n)$ for a constant $C$ and for all $n\in\mathbb{N}$.

   \item[(d)] {\it Error bounds} (cf. Theorem 5.4.4 in ref.~\cite{BGM} and Theorem 16.2 in ref.~\cite{Baker1975Essentials}.) 
   Padé approximants $P^{M+J}_{M}(z;x_0)$ are good representations of the function $f(z)$ at points that are not too close to the branch cut. To make this precise, consider $\Delta>0$ and define the region
\begin{equation}
    \mathcal{D}^{+}(\Delta)= \{x+i~ y \in \mathbb{C}~|~ x \le -R, ~|y| \ge \Delta\}~ \cup~ \{x+i~ y \in \mathbb{C}~|~ x> -R\}\,,
\end{equation} 
as shown in Fig.~\ref{fig:pade}.
   If the asymptotic expansion of $f(z)$ given in eq. (\ref{asyexp}) is convergent, 
   then for any point $z \in \mathcal{D}^{+}(\Delta)$, 
   the absolute difference between the Padé approximants and the function is bounded as follows,
\begin{align}\label{eq:Stieltjes-Pade-bound}
       |f(z)- P^{M+J}_{M}(z;x_0)| < c \bigg|\frac{(z-x_0)}{\rho}\bigg|^{J+1} \Big| \tfrac{\sqrt{\rho+z-x_0}-\sqrt{\rho}}{\sqrt{\rho+z-x_0}+\sqrt{\rho}} \Big|^{2M},\quad \forall J\ge -1,M \ge 1 \,,
   \end{align}
where $\rho= R+x_0-\Delta$ and $c$ is a constant.
\end{itemize}

In summary, Stieltjes functions have excellent analytic properties. They can be described  by Padé approximants, both along the real axis and in the cut complex plane. Moreover, these approximations come with rigorous error bounds. One of the major advantages of the method is that, once computed, the Padé approximants can be stored. Consequently, once the Padé approximants have been obtained, the method becomes highly efficient for numerical evaluation over a large sample of phase-space points, since this reduces to evaluating stored rational functions at the required points. In subsection \ref{sec:subsecPadeexample}, we illustrate this with an example relevant to five-particle scattering amplitudes.

\subsection{Proof of Stieltjes property for a class of Feynman integrals}
\label{sec:FeynStielt}

We now show that, subject to certain criteria, Feynman integrals are Stieltjes functions.
Here, we restrict our attention to scalar Feynman integrals. 
Our proof uses their standard Feynman parametrization. We recall its explicit form in Appendix~\ref{sec:Feynman}, cf. eq. (\ref{eq:FeynmanreprAppendix}). The key feature relevant to this discussion is that this formula has the following schematic form (dropping irrelevant prefactors),
\begin{align}\label{eq:Feynmanschematic0}
I(\{x_i\}) \sim \int_0^{\infty}  \frac{\prod_i  d\alpha_i~\alpha_i^{\nu_i-1}}{\rm{GL}(1)}  \frac{U^{q}}{F^{e}} \,,
\end{align}
with the second Symanzik polynomial given by
\begin{align}\label{eq:Symanzik2}
    F = \sum_{i} x_i A_i \,,
\end{align}
and with $U$ and $A_i$ being non-negative homogeneous polynomials in the integration variables $\alpha_i$.
The exponents $\nu, q, e$ are subject to convergence conditions.
The variables $x_i$ are given by the set of (negative) Mandelstam invariants and squared masses, see Appendix~\ref{sec:Feynman}.

In the Euclidean region, $F \ge 0$ by definition. In order to prove the Stieltjes property with respect to a given variable $x$, we need a further positivity property with respect to the coefficients of $x$ in $F$, which is
\begin{align}\label{eq:technicalcondition}
F = A \, x + B \,, \quad {\rm with} \quad A \ge 0\,, B \ge 0\,,
\end{align}
where the second inequality is understood with the kinematic variables taking values within the Euclidean region.

This is immediate to show for planar Feynman graphs. 
In that case, the polynomials $A_i$ corresponding to non-planar kinematic variables vanish. In other words, apart from the squared masses, only the planar variables $-(p_m + p_{m+1} + \ldots + p_{n-1})^2$ appear as $x_i$ in eq. (\ref{eq:Symanzik2}).
Let us single out one of these variables, $x=x_i$, and write $A_i = A, B = \sum_{i \neq j} x_i A_i$.
We then have eq. (\ref{eq:technicalcondition}), together with the required positivity conditions.

Let us now comment on possible extensions of this property to the non-planar case.
Here a subtlety arises, which is that the set of $x_i$ comprises both planar and non-planar variables that are not independent. For example, for massless four-particle scattering, it contains the variables $-s,-t,-u, -p_1^2 , -p_2^2 , -p_3^2, -p_4^2$, which satisfy
the constraint 
\begin{align}\label{eq:constraintfourparticle}
    s+t+u = p_1^2 +p_2^2 + p_3^2+p_4^2 \,.
\end{align}
This means that only six of the seven variables are independent, and eliminating one of them may introduce minus signs. 
One possibility to proceed is relax the constraint (\ref{eq:constraintfourparticle}), i.e. to take eqs. (\ref{eq:Feynmanschematic0}) and (\ref{eq:Symanzik2}) as the definition of a generalized Feynman diagram, as in ref. \cite{Tausk:1999vh}. The latter diagram has the desired Stieltjes property in all variables $x_i$ (as we show presently). One could then benefit from this property, and impose the constraints, such as (\ref{eq:constraintfourparticle}), at the end of the calculation. 
Alternatively, one may look for a choice of independent variables, such that eq. (\ref{eq:technicalcondition}) holds.
A possible general proof that this can be achieved goes beyond the scope of this work. However, we provide evidence that this is possible in non-trivial examples of non-planar two- and three-loop Feynman integrals, in Appendix \ref{sec:Feynman}.

Provided that the technical condition (\ref{eq:technicalcondition}) holds, we have
\begin{align}\label{eq:Feynmanschematic}
I(x) \sim \int_0^{\infty}  \frac{\prod_i  d\alpha_i~\alpha_i^{\nu_i-1}}{\rm{GL}(1)}  \frac{U^{q}}{\left(A \,x\,+B \right)^{e}} \,,
\end{align}
with $A \ge 0$ and also $B\ge 0$, assuming that the $\{ x_i \}$ are set to values in the Euclidean region.
We now argue that the Stieltjes property follows from the structure of the Feynman parametrization in eq.~\eqref{eq:Feynmanschematic}.

\begin{enumerate}
    \item \textit{Analyticity:} 
    The integral $I(x)$ in eq.~\eqref{eq:Feynmanschematic} can develop a singularity only if the integrand itself becomes singular. For fixed $x$, with all other variables taken in the Euclidean region, this is possible only when $x$ lies on the negative real axis. This implies that the function is analytic in the cut plane 
    $\mathbb{C}\backslash(-\infty,-R]$, with some $R \ge 0$.
    
 \item \textit{Asymptotic behavior:} 
Let $x=r~ e^{i\theta}$ with $|\theta|<\pi$. Using 
\begin{align}
|A~ r~ e^{i\theta}+B|\ge C_\theta |A~r+B|, \qquad 
C_\theta=\begin{cases}
\sin(\tfrac{\theta}{2}), & \theta\neq0\\
1,&\theta=0
\end{cases}\,,
\end{align}
we see that the absolute value of the integrand is dominated by a function proportional to $(A\, r + B)^{-e}$, which is integrable because the Feynman integral is known to converge for positive real $x$. 
Since the integrand vanishes pointwise in the limit of large $r$, Lebesgue's  theorem applies and ensures that $\lim\limits_{r\to\infty} I(r e^{i\theta})=0$.

\item \textit{Herglotz property:}~
 Starting again from the integrand of eq. (\ref{eq:Feynmanschematic}), we distinguish the following two cases:
\begin{itemize}
\item \textit{Case (1): $e = 1$.}  Using
\begin{align}
\frac{1}{A x + B} 
 = \frac{A x^{*} + B}{|A x + B|^{2}} \,,
\end{align}
one verifies eq. (\ref{eq:herglotz}), and therefore $I(x)$ is a Stieltjes function.
\item \textit{Case (2): $0 < e < 1$.}   This is relevant for example in dimensional regularization. We employ the identity
\begin{align}
 \frac{1}{(A x+B)^{e}} =    \frac{\sin{\pi e}}{\pi}\int\limits_{0}^{\tfrac{A}{B}} \frac{y^{e-1}}{1+ x~ y}~ \frac{1}{(A-B y)^e} dy \,, \quad \quad 0<e<1.
\end{align}
Comparing this to eq. \eqref{eq:Stieltjesdefinition}, we see that $(A x+B)^{-e}$ is a Stieltjes function for $0<e<1$, with $ A>0, B>0$. Since the space of Stieltjes functions is convex \cite{Bender}, the subsequent integrations over the Feynman parameters in eq. (\ref{eq:Feynmanschematic}) preserve this property. Hence, $I(x)$ is a Stieltjes function.
 \end{itemize}
\end{enumerate}
This concludes the proof of the Stieltjes property for planar Feynman integrals, subject to the condition $0<e \le 1$.

\begin{figure}[t]
    \centering
    \subfloat[]{\includegraphics[width=0.3\textwidth]{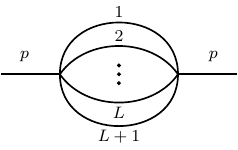}}\hfill
    \subfloat[]{\raisebox{1.06mm} {\includegraphics[width=0.3\textwidth]{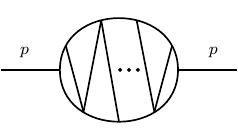}}}\hfill
    \subfloat[]{\raisebox{3.6mm}{\includegraphics[width=0.3\textwidth]{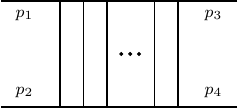}}}\\
    \caption{Examples of Feynman integrals that are Stieltjes. (a) `Banana' integrals in $D=2$. (b) Zig-zag integrals in $D=4$. (c) Ladder integrals in $D=6.$}
    \label{fig:Stieltjes_pdf}
\end{figure}

 In summary, we have proved the following theorem.
\begin{tcolorbox}[arc=3mm]
{\bf Theorem [Sufficient condition for scalar Feynman integrals to be Stieltjes functions]:}~Scalar Feynman integrals, 
as defined in eq.~(\ref{eq:FeynmanreprAppendix}), 
are Stieltjes functions, provided that
$0<\sum_{i} \nu_i - L D/2\le 1$, where $L$ is the loop order, $D$ is the space-time dimension, and $\nu_i$ are the propagator powers, and provided that a technical condition holds, namely that the second Symanzik polynomial can be put into the form $F =  A \,x+ B$, where $A\ge 0, B\ge 0$.
The Stieltjes property then holds in the Euclidean region.
This holds for general planar Feynman integrals, where $x$ can be chosen to be any of the variables $-(p_i + p_{i+1} + \ldots + p_{j-1})^2$ or $m_{i}^2$, with the other variables held fixed in the Euclidean region.
In the case of non-planar Feynman integrals, we have provided evidence in the form of examples that the technical condition holds, but a proof is left to future work.
\end{tcolorbox}
This generalizes the complete monotonicity (CM) property proven in ref.~\cite{Henn:2024qwe}. 
See Fig.~\ref{fig:Stieltjes_pdf} for examples of Feynman integrals that saturate the upper bound of the inequality. 
We emphasize that our proof provides a sufficient criterion for a Feynman integral to be a Stieltjes function, and not a necessary condition.

We would like to comment that the choice of variables to observe Stieltjes properties in is not unique, even in the planar case. It would be interesting if the methods of \cite{Sturmfels:2025wpg} could be extended to shed more light on possible variable choices, and possibly to prove the technical condition (\ref{eq:technicalcondition}) in the non-planar case.
Related to this, it would be interesting to extend our framework from treating one variable at a time to genuine multi-variable approaches. For an overview of related multi-variate Padé approximants, see
\cite{Cuyt:1983}. For multi-variable extensions of complete monotonicity, cf. refs.~\cite{Henn:2024qwe,Mazzucchelli:2025gyg}.

\subsection{Example relevant to one-loop five-particle scattering}
\label{sec:subsecPadeexample}

As an example of the relevance of Stieltjes functions and their usefulness, we discuss functions arising in one-loop five-particle scattering processes, as studied in ref.~\cite{Bern:1993mq}. 
In that work, the amplitudes were expressed in terms of logarithms and dilogarithms. 
The authors of ref.~\cite{Bern:1993mq} identified certain special combinations of these functions that have nice properties, e.g. with respect to their singularity structure. 
We demonstrate that these combinations are closely related to Stieltjes functions.
Let us consider the following functions,
\begin{align}
f^{(0)}(x)=& \frac{1}{x} \,,\\
f^{(1)}(x)=& \frac{\log{x}}{x-1} \,, \label{eq:examplelog}\\
f^{(2)}(x,y) =&  \frac{1}{1-x-y} \left[ {\rm Li}_{2}(1-x) + {\rm Li}_{2}(1-y)  + \log x \log y -\frac{\pi^2}{6} \right] \label{eq:exampleLi2} \,.  
\end{align}
One can show that these are all Stieltjes functions, and hence
they enjoy the properties discussed in section \ref{sec:stieltjes}. 
It can then be readily verified that all transcendental functions appearing in ref.~\cite{Bern:1993mq}, in particular those appearing in eq. (5) of that work, can be expressed in terms of simple $\mathbb{Q}$-linear combinations of $f^{(i)}$ and their derivatives.
Let us now focus on $f^{(1)}(z)$ defined in eq. (\ref{eq:examplelog}), but for a complex argument $z$ in the cut plane, to illustrate the convergence properties of Padé approximations, discussed in section \ref{sec:subsecPadeconvergence}. To see that $f^{(1)}(z)$ is a Stieltjes function, consider its (dispersive) representation,
\begin{equation}
  \frac{\log{z}}{z-1} = \int_{0}^{\infty} du~ \frac{1}{1+u ~z}~\frac{1}{(1+u)}\,. 
\end{equation}
This corresponds to the integral representation defined in eq. \eqref{eq:Stieltjesdefinition}, with $\rho(u)=1/(1+u)$ and $a=0,~b=\infty$. The Taylor expansion of the function  $f^{(1)}(z)$ around $x_0=1$ is given by
\begin{equation}\label{eq:logxbx-1exp}
  f^{(1)}(z;1) = \sum\limits_{n=0}^{\infty} \frac{(-1)^n}{n+1} (z-1)^n   \,,
\end{equation}
with a radius of convergence $R=1$. By truncating eq.~\eqref{eq:logxbx-1exp} to the first $2 N$ terms, one can construct the Padé approximations $P_N^{N-1}(z)$ and $P^N_{N}(z)$.
In practice, this can be done using the {\tt Mathematica} command {\tt PadeApproximant}, for example. The first non-trivial case, i.e. $N=1$, is given by
\begin{equation}
\begin{aligned}
P_{1}^{0}(z;1)=\frac{2}{1+z}\,, \quad\text{and}\quad    P_{1}^{1}(z;1)=&\frac{z+5}{2 (2 z+1)}\,. 
\end{aligned}
\end{equation}
\begin{figure}[t]
    \centering
    \subfloat[$N=1;~M=1$\label{fig:log11}]{\includegraphics[width=0.45\textwidth]{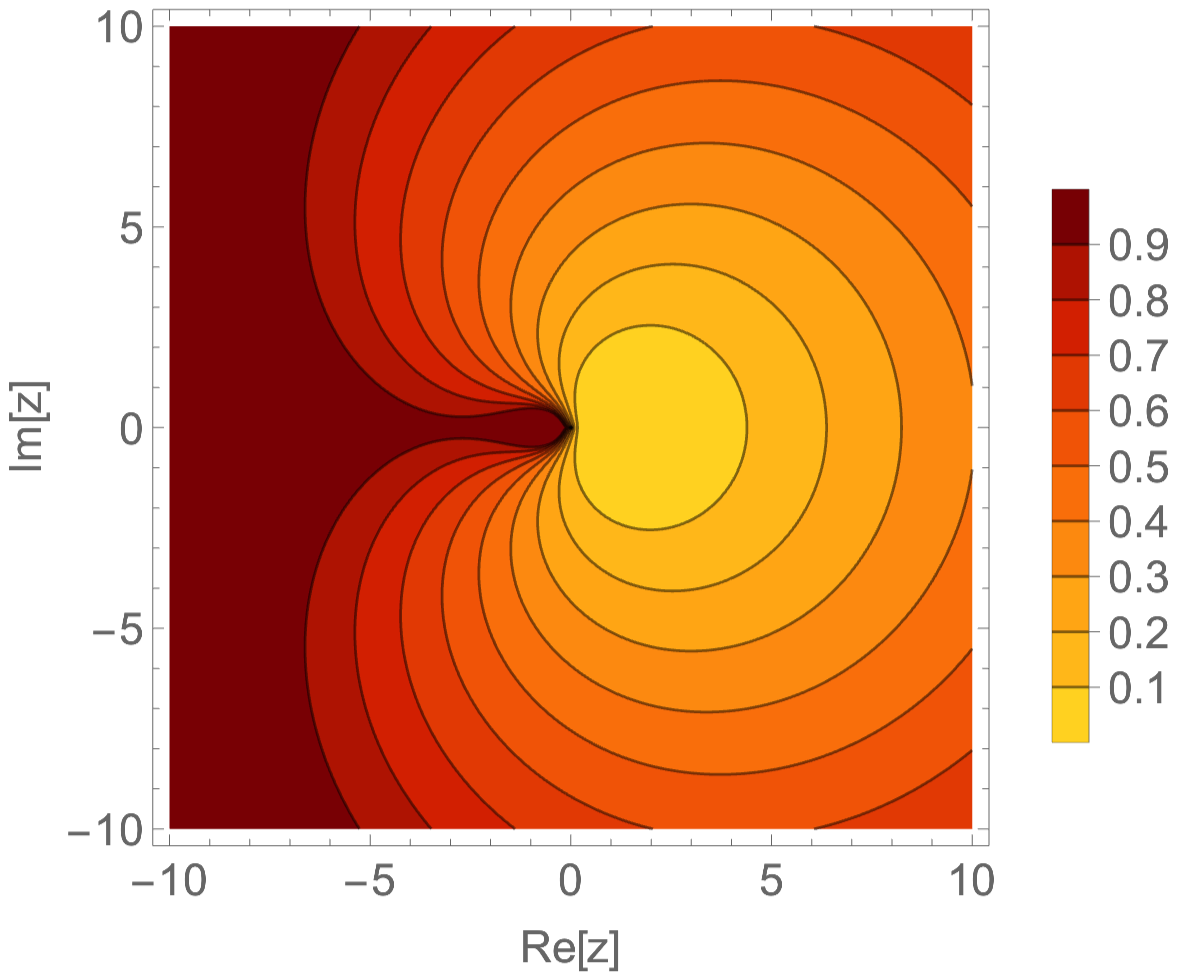}}\hfill
    \subfloat[$N=0;~M=1$\label{fig:log12}]{\includegraphics[width=0.45\textwidth]{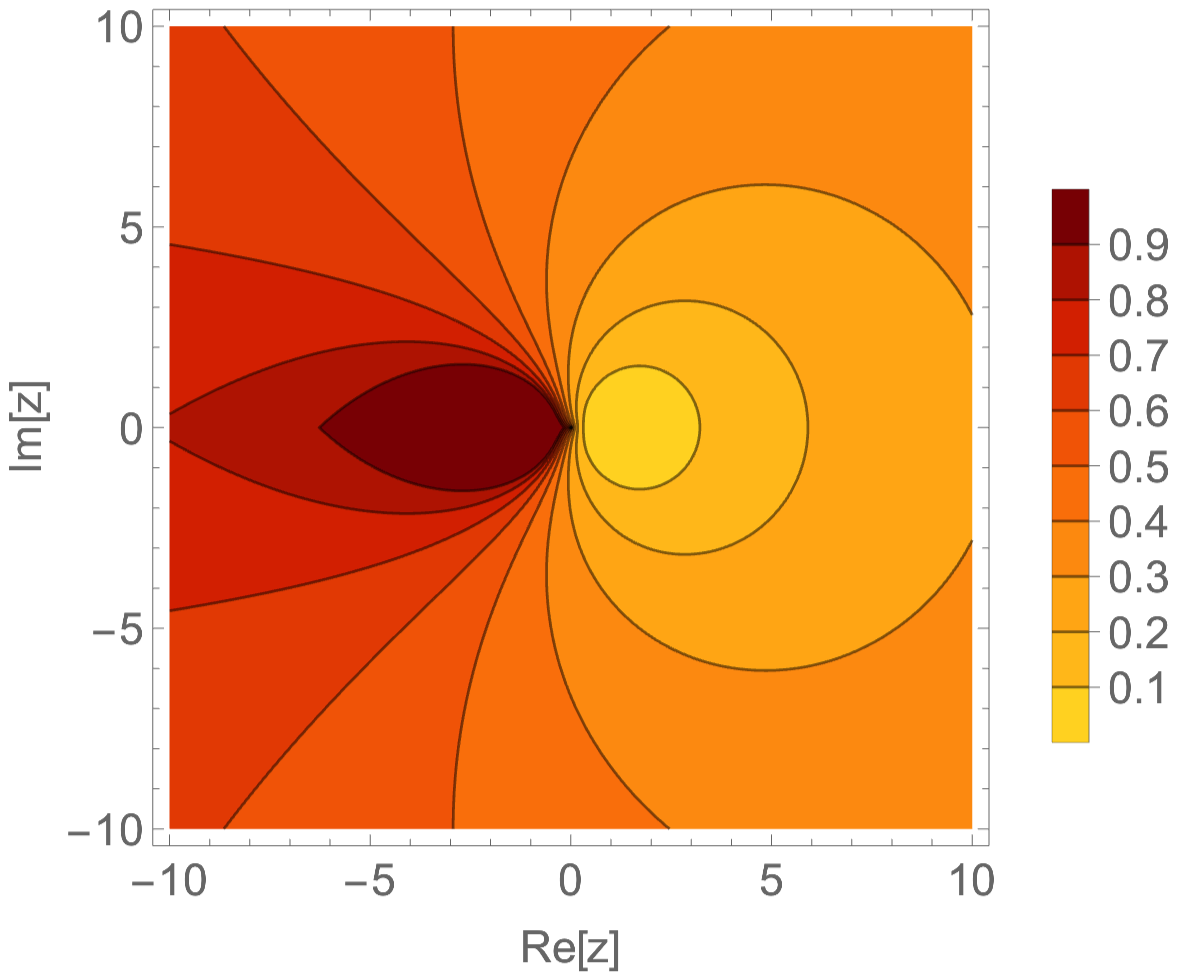}}\\
    \subfloat[$N=10;~M=10$\label{fig:log1010}]{\includegraphics[width=0.45\textwidth]{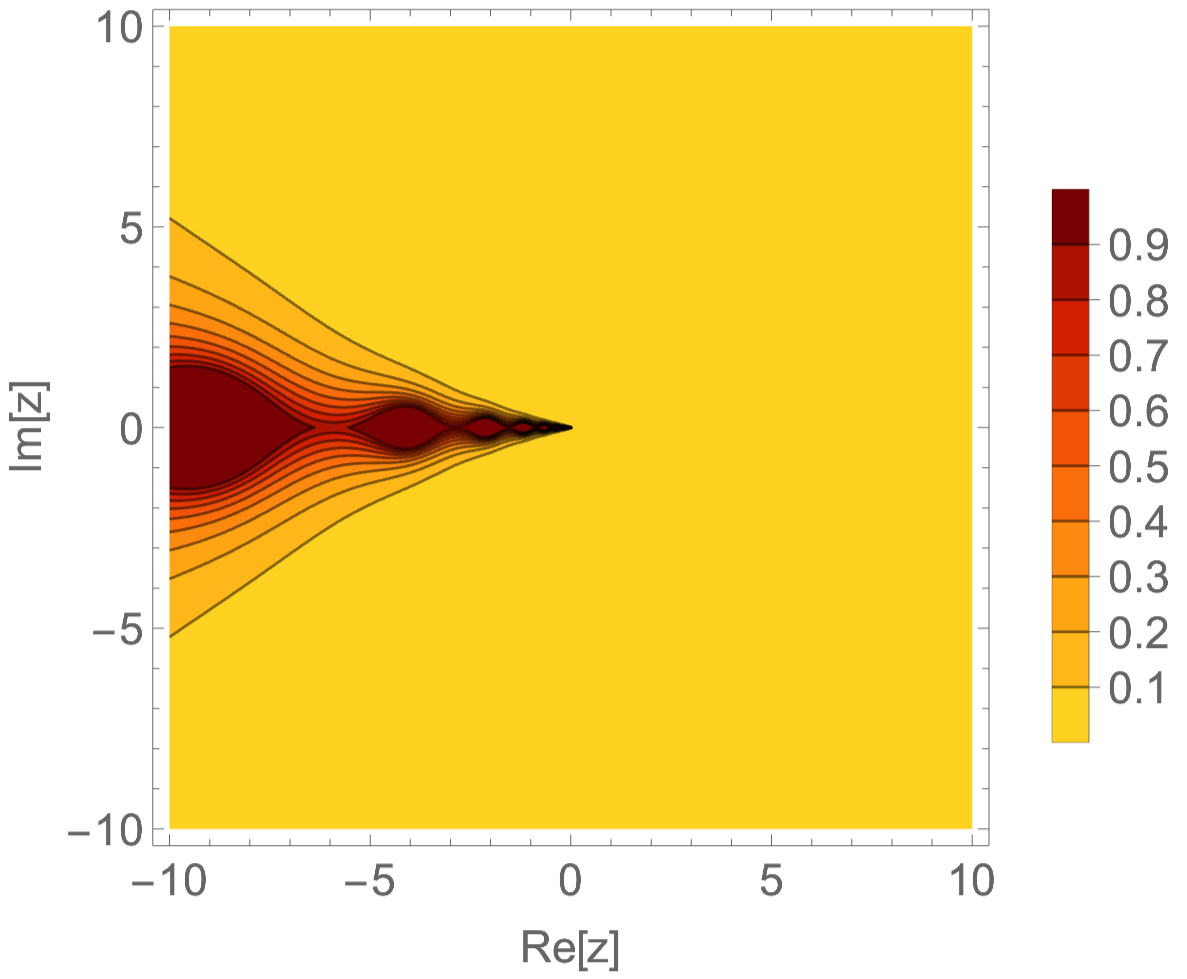}}\hfill
    \subfloat[$N=9;~M=10$\label{fig:log1011}]{\includegraphics[width=0.45\textwidth]{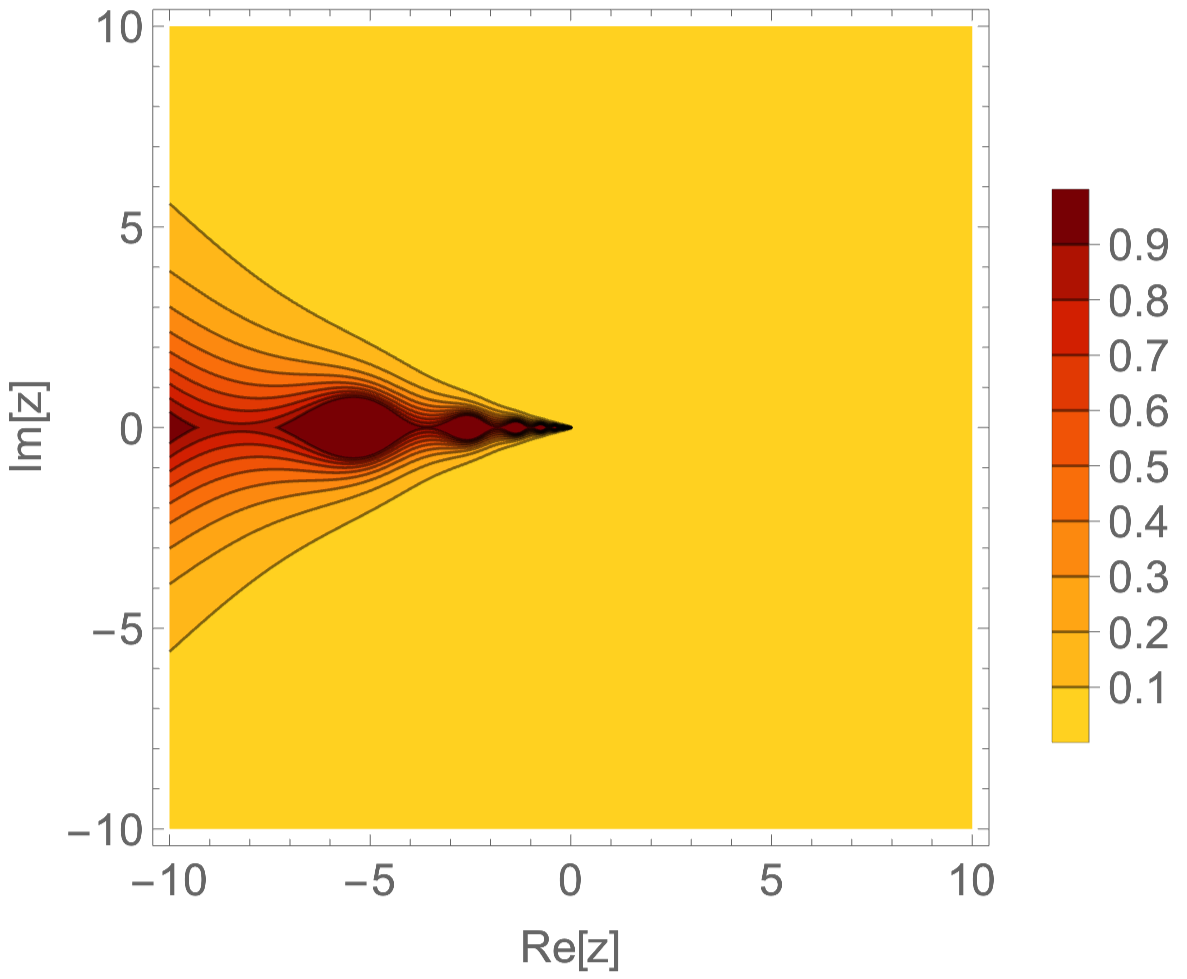}}\\
    \caption{The relative error $r(z)$ of the Padé approximant $P_{N}^{M}(z;1)$ with respect to the function $f^{(1)}(z)$ in the complex plane, for different choices of $N$ and $M$. The red areas indicate a relative error of 100\% or more. The Padé approximant is a relatively good approximation already for small $N$, as can be seen in panels (a) and (b).
        We see that the approximation is best close to the positive real axis, and worst the negative real axis, as expected. Panels (c) and (d) show a significantly increased precision.}
    \label{fig:relerr}
\end{figure}
We can now convince ourselves that the properties of the Padé approximants discussed in section \ref{sec:subsecPadeconvergence} indeed hold. We start by demonstrating convergence on the real axis. For any real \(z \in \mathbb{R} \setminus (-\infty,0)\), eq.~\eqref{realx} is satisfied. In other words, the Padé approximants $ \{P^{N-1}_{N}(x;1) \}$ and  $ \{P^{N}_N (x;1)\}$  provide a lower and upper bound to $f^{(1)}(z)$, respectively. For example, for $N=1$, eq.~\eqref{realx} implies
 \begin{align}
    \frac{2}{1+z}\le \frac{\log{z}}{z-1} \le \frac{z+5}{2 (2 z+1)}\quad \forall z\ge 1\,.
 \end{align}One can also observe the convergence in the cut plane, as discussed in point (c) of section \ref{sec:subsecPadeconvergence}. In Fig.~\ref{fig:relerr}, we show the relative error of the Padé approximants $P_{N}^{M}$, i.e.,
 \begin{equation}
     r(z)=\Bigg|\frac{P_{N}^{M}(z;1)-f^{(1)}(z)}{f^{(1)}(z)}\Bigg| \,,
 \end{equation}
  for different values of $N$ and $M$. We find, that the upper and lower bound are close to the function and that the error is biggest close to the branch cut. We also see that the Padé approximants converge fast numerically when increasing $N$ and $M$. 
  
  The convergence also holds for expansion around any complex value \(z=z_0 \in \mathcal{D}^{+}(1/100)\) and for any desired precision. Take for example $z_0 = 23 + 45 i$ and $20$ decimal places. We can use eq.~\eqref{eq:Stieltjes-Pade-bound}  
  to determine a value of \(M\) above which \(P_M^{M+J}(z;1)\) is guaranteed to converge for fixed $J$. 
 In this case, for \(J = -1,0\), we find \(M = 97\) and \(M=105\), respectively.

\subsection{Padé-powered analytic continuation of CM boostrap results}
\label{sec:PadepowerCM}

It is natural to combine the CM bootstrap of section~\ref{sec:CMbootstrap} 
with the insights about Stieltjes functions.
One can use the CM bootstrap to provide high-precision numerical values in a kinematic region. Then, the Padé approximation is determined and used to transport the numerical values to other kinematic regions. For Stieltjes functions, this is efficient thanks to the convergence proofs discussed above. For these reasons, we consider this as an attractive alternative to traditional integration methods of differential equations.
In detail, we proceed according to the following steps:
\begin{itemize}
\item {\it Step~1: Select a starting point.}
Identify  the regions in which two-sided bounds can be obtained, and choose a point $x_0$ within this two-sided region with good convergence. 

\item {\it Step~2: Compute the basis integrals at the starting point.} Estimate the basis integrals ${\bf f}(x_0)$ at $x_0$, using the complete monotonicity bootstrap. Choose the number of derivatives $n_0$ in the CM bootstrap needed to achieve a given precision, say of $k$ relevant digits, of the basis integrals. 
 
\item {\it Step~3: Compute the Taylor expansion.} Rationalize ${\bf f}(x_0)$ for numerical stability and compute the Taylor expansion of the basis integrals around $x_0$ to $n$ terms. Using the recursion relation for the derivatives obtained from the differential equation, cf. eq.~\eqref{eq:recursiondericative}, the Taylor series takes the form 
    \begin{equation}\label{eq:taylorexp}
        {\bf f}(z;x_0)
        = \sum_{i=0}^{n} 
        (-1)^i\frac{ Q_i(x_0)  {\bf f(x_0)}}{i!}\,
        (z - x_0)^i +\mathcal{O}\big((z-x_0)^{n+1}\big)\, .
    \end{equation}
Since our basis integerals were estimated to a precision of $k$-digits and the entries of the differential equation matrix are rational, we can ensure that the Taylor coefficients are also accurate to $k$-digits. 
\item {\it Step~4: Construct the Padé approximants.}
Find the Padé approximants of eq.~\eqref{eq:taylorexp} around $x_0$, to obtain $P^{N-1}_{N}(z;x_0)$ and $P_N^{N}(z;x_0)$, with $2N \le n$. Once obtained, store these rational functions. 

    \item {\it Step~5: Evaluate the Padé approximants.} Evaluate the Padé approximants at any desired point $z = z_0$ in the cut plane $\mathcal{D}^{+}(\Delta)$. This gives an upper and lower bound on the function value at real $x = x_0$, as given in eq. \eqref{realx}, and approximations of the function value according to eq. \eqref{eq:Stieltjes-Pade-bound} for complex $z=z_0$, allowing for an efficient numerical evaluation.
\end{itemize}
One can increase $n_0$, and therefore $k$, or $N$, which might require increasing $n$, and repeat the above steps if needed until one obtains results of the required precision. The precision achieved can be checked, for real values, thanks to the bounds given in eq.~(\ref{realx}).

\begin{figure}[t]
    \centering
    \includegraphics[width=0.9\linewidth]{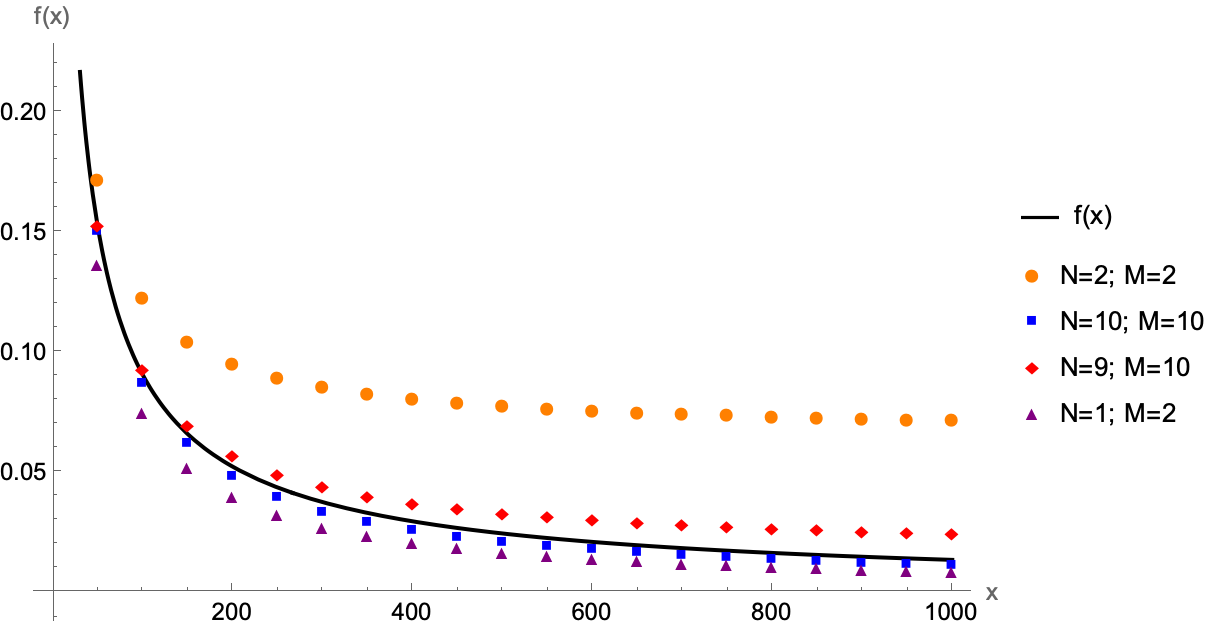}
    \caption{The Padé approximations around $x_0= -1/10$ compared with the 
    function for different $N$ and the massive bubble integral for positive real $x$.
 }
    \label{fig:1a}
\end{figure}

This method has the advantage that we only have to compute one starting point, which we can choose so that the evaluation with the CM bootstrap is particularly efficient. One can then compute the Padé approximants around this starting point once and store them. These rational functions can subsequently be evaluated at many phase space points with high precision and efficiency. 

Let us illustrate this in detail on the massive bubble example. 
The latter is a Stieltjes function in $D=2$, as shown in section \ref{sec:FeynStielt}. 
The interval with two-sided bound is $(-2,0)$ in this case, and we know that convergence is best close to the upper end, which is why we chose $x_0=-1/10$. 

Let us choose $k=10$ digit precision for the Taylor coefficients. Using the CM bootstrap with up to $n_0=20$ derivatives, we obtain $f(x_0=-1/10)=1.017007305$. We can then compute the Taylor series by inserting $x_0$ and $f(-1/10)$ into eq.~\eqref{eq:taylorexp}.
From this, we compute the two Padé approximants, which provide an upper and a lower bound according to eq.~\eqref{realx}. For $N_{\text{max}}=n/2=2$, the highest Padé approximants we can compute are
$P^1_2(z;-1/10)$ and $P^2_2(z;-1/10)$. We can evaluate these at different $z=z_0$ to obtain numerical values of the function $f(z)$. For example, for $z_0=1.5$, we find $P^1_2(z_0;-1/10)=0.8068937$ and $P^2_2(z_0;-1/10)=0.8068920$. This is to be compareed with the function value $f(z_0)=0.8068922$. Hence, the inequality in eq.~\eqref{realx} is satisfied. As one can see, the constraints are already strong, even though we included only $n=4$ terms in the Taylor expansion. 

\begin{figure}[t]
    \centering
    \includegraphics[width=0.6\linewidth]{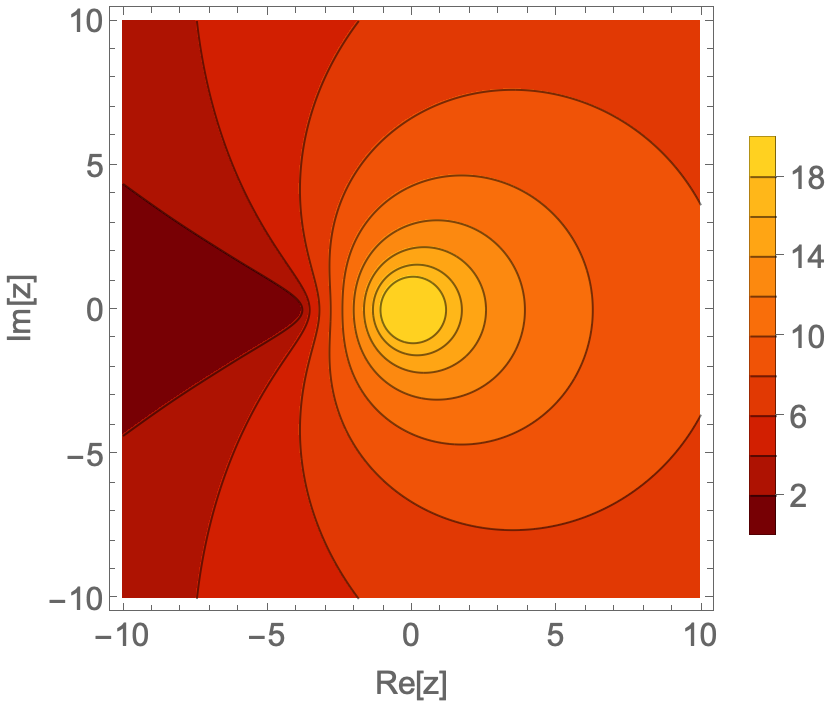}
    \caption{Agreement in numbers of digits between the Padé approximant $P_{10}^{10}(z;x_0)$ around $x_0= -\frac{1}{10}$ and the massive bubble integral $f(z)$, see eq. \eqref{relerrbubble}. The light yellow area indicates an agreement of at least 18 digits. The dark red area indicates an agreement of at most 2 digits.}
    \label{fig:1b}
\end{figure}

In Fig. \ref{fig:1a}, we show the convergence of the Padé approximants when increasing $N_{\text{max}}$. We see that the upper and lower bound converge closer to the function value if we choose $N_{\text{max}}=10$ instead of $N_{\text{max}}=2$. While the approximation is very good for small values of $z_0$ already for $N_{\text{max}}=2$, $P^{10}_{10}(z_0;-1/10)$ and  $P^{9}_{10}(z_0;-1/10)$ approximate the function reasonably well even for values of $z_0$ as large as $1000$, with a relative error of around 6\% in that case.

The convergence for complex $z_0$ is also good. For instance, if we choose $z_0=2+3i$, we obtain $P^1_2(z_0)=0.6724847 - 0.2231611 i$, $P^2_2(z_0)=0.6715767 - 0.2225451 i$, to be compared to $f(z_0)=0.6717457 - 0.2224894 i$. In Fig.~\ref{fig:1b}, we show the negative logarithm of the relative error between the Padé approximant $P_{10}^{10}(z;x_0)$ and the function in the complex plane,
\begin{equation}
\label{relerrbubble}
    d(z;x_0)= - \text{log}_{10}\Bigg(\bigg| \frac{f(z)-P_{10}^{9}(z;x_0)}{f(z)}\bigg|\Bigg)\,.
\end{equation}
This quantity measures the number of digits of agreement. As one can see, we obtain very good agreement close to the expansion point, and worse agreement for values approaching the branch cut.

\begin{figure}
    \centering
    \includegraphics[width=0.6\linewidth]{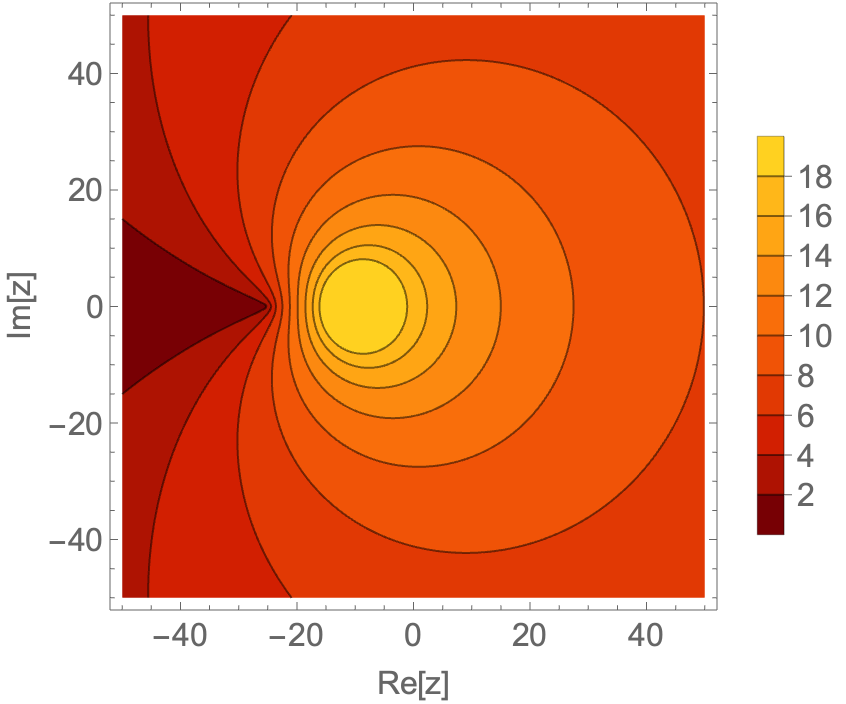}
    \caption{Precision, i.e. number of digits obtained, of the Padé approximant for $N=50$ of the four-loop `banana' integral. The light yellow area indicates an agreement of at least 18 digits. The dark red area indicates an agreement of at most 2 digits.}
    \label{fig:pade4l}
\end{figure}

The multi-loop `banana' integrals with unit propagator powers are Stieltjes, as shown in section \ref{sec:FeynStielt}. For those integrals, we compute the Padé approximants. This takes a couple of seconds only, even including a high number of derivatives. We chose a starting value close to the upper bound of the region with two-sided bounds, where the convergence is very quick already for a low number of derivatives. The  Padé approximants can be stored and evaluated in the cut complex plane, which again is very fast as they are just rational functions of the argument $z$. 

For example, for the four-loop `banana' integral, we choose the starting point $x_0=-91/10$ and include $n_0=100$ derivatives, which allows us to get Taylor coefficients with a precision of $k=50$ digits. Then evaluating the integral at the starting point using the CM bootstrap takes 22\,s and computing the Taylor expansion and the Padé approximants at $N=50$ takes 23\,s. Once we have the Padé approximants, we can evaluate them anywhere in the cut plane. In Fig. \ref{fig:pade4l} we show the precision, i.e. the number of digits that can be obtained from this setup, by plotting 
\begin{equation}
\label{prec}
    d(z;x_0)= - \text{log}_{10}\Bigg(\bigg| \frac{P_{50}^{50}(z;x_0)-P_{50}^{49}(z;x_0)}{\frac{1}{2}\big(P_{50}^{50}(z;x_0)+P_{50}^{49}(z;x_0)\big)}\bigg|\Bigg)\,,
\end{equation}
in the cut complex plane.
(This is a reasonable measure,
since both $P_{49}^{50}(z;x_0)$ and $P_{50}^{50}(z;x_0)$ are expected to converge to the function.)
We find that the precision is best close to the expansion point and decreases slowly away from it. Again, the precision is worst close to the branch cut on the negative real axis (which starts at $x=-25$ in this case). 
Let us also comment on the runtime. As the method relies just on the evaluation of rational functions, we expect little dependence on the phase space point.
We randomly pick ten points and compute the average evaluation time per point. We find that this takes 0.22\,s per point only. This is almost two orders of magnitude better than the time we found for \texttt{AMFlow}, see Table \ref{PerformanceComparison}.

\subsection{Numerical evaluation of a 20-loop `banana' integral}
\label{sec:twenty-loop-banana}

We wish to stress that the insights on Stieltjes properties of Feynman integrals can be used independently of the CM bootstrap. 
It does not require the knowledge of a differential equation. 
In practical situations, it is often possible to obtain expansions in a different way, e.g. by expanding around a special point.

Let us illustrate the power of this approach on the example of the 20-loop equal-mass `banana' integral.
\begin{table}[t]
\renewcommand{\arraystretch}{1.3}
\begin{center}
\begin{tabular}{c|c|c|c}
$z$& $P_{10}^{9}(z;0)$ & $I_{20}(z)$ & $P_{10}^{10}(z;0)$ \\
\hline
 $10^3$ & $1.53585179150421\times 10^{13}$ & $1.53585179150421\times 10^{13}$ & $1.53585179150421\times 10^{13}$ \\
 $10^4$ & $1.534895905\times 10^{13}$ & $1.534895909\times 10^{13}$ & $1.534895911\times 10^{13}$ \\
 $10^5$ & $1.52772\times 10^{13}$ & $1.52785\times 10^{13}$ & $1.52797\times 10^{13}$ \\
\end{tabular}
\end{center}
\caption{Comparison of the Pad\'e approximants with the Bessel integral representation of the 20-loop `banana' integral in $D=2$.}
    \label{tab:4}
\end{table}The equal mass $L$-loop `banana' integral $I_L(x)$ in $D=2$ admits a representation as a one-fold Bessel integral in the Euclidean region, i.e. for $x \ge -(L+1)^2$,
see refs.~\cite{Groote:2005ay,Pogel:2022vat},
\begin{equation}\label{eq:Besselint}
 I_L(x)= 2^{L}~\int_{0}^{\infty} dt~ t~J_{0}(t \sqrt{x}) K_0(t)^{L+1}\,,
\end{equation}
where $J_0(z)$ and $K_0(z)$ are the Bessel function of the first kind and the modified Bessel function of the second kind, respectively. The Bessel integral representation can be used to derive the Picard–Fuchs differential equations satisfied by these integrals to very high loop orders; this has been carried out up to $L=20$, see refs.~\cite{vanhove2014physicsmixedhodgestructure, delacruz2024algorithmdifferentialequationsfeynman}.

We use eq. (\ref{eq:Besselint}) to obtain the Taylor expansion around the origin,
\begin{equation}\label{eq:Taylor}
    f(x)= \sum_{n=0}^{\infty}(-1)^n a_n x^n  \,,
\end{equation}
with the moments given by 
\begin{equation} \label{eq:Besselmom}
 a_n = \frac{4^{-n}}{n!^2}\int_{0}^{\infty} dt~ t^{2n+1} K_0(t)^{L+1}\,.
\end{equation}
We have omitted the overall factor of $2^L$ from our analysis to keep the numbers obtained manageable. These Bessel moments 
have been computed
in closed form for $L=1,2,3$, see ref.~\cite{Bailey:2008ib}. 
To achieve a reasonable precision for the $20$-loop `banana' integral, we compute numerically the first $20$ Bessel moments in eq. \eqref{eq:Besselmom}, to an accuracy of $50$ digits i.e., $n=20$ and $k=50$. This takes roughly $35$\,s. We then use this information to obtain the corresponding Padé approximants. 

\begin{figure}[t]
    \centering
   \includegraphics[width=0.6\linewidth]{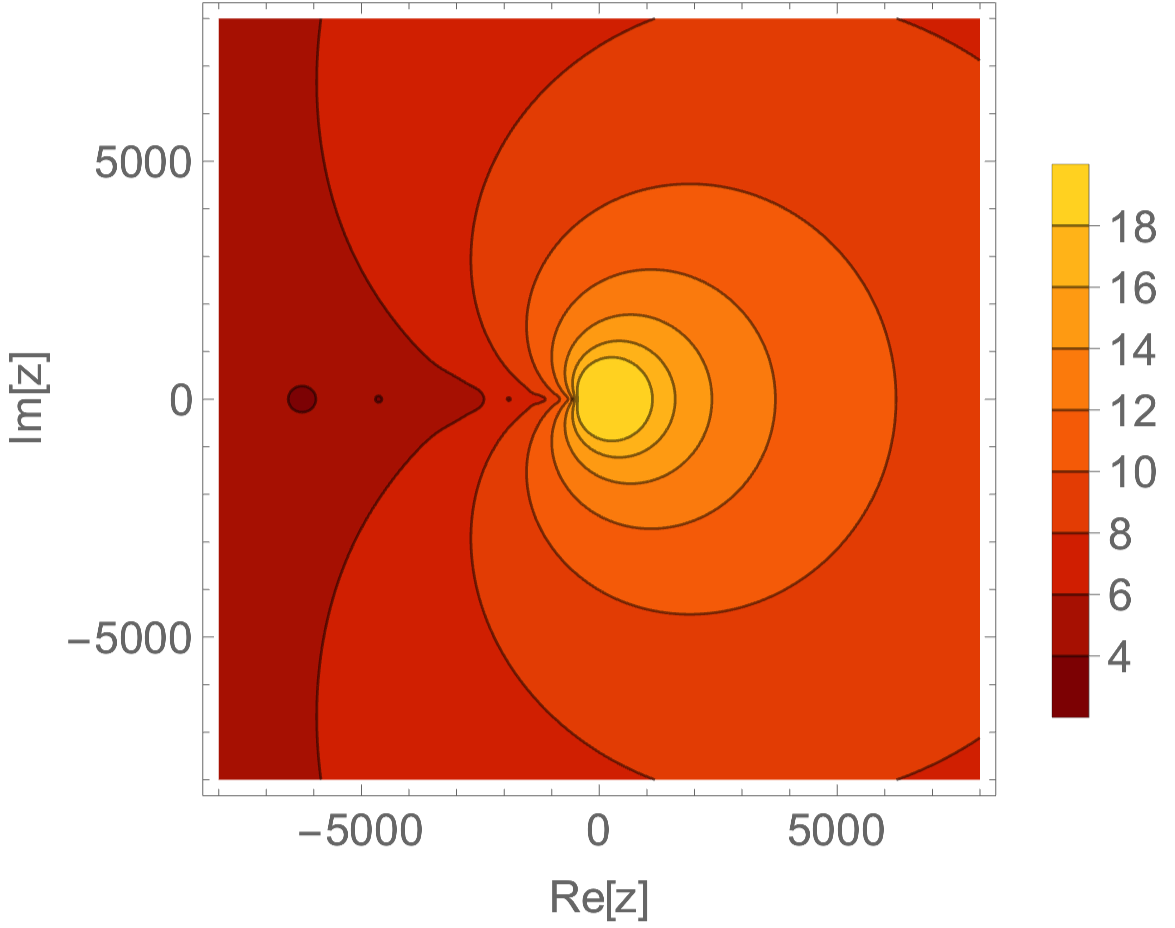}
    \caption{Precision, i.e. digits obtained, of the Padé approximant for $N=10$ of the 20-loop `banana' integral. The light yellow area indicates an agreement of at least 18 digits. The dark red area indicates an agreement of at most 4 digits.}\label{fig:20_L_banana}
\end{figure}

The result is shown in Tab. \ref{tab:4}. In that table, we compare the Padé approximants $P_{10}^{9}(z;0)$ and $P_{10}^{10}(z;0)$ with the value obtained from eq.~\ref{eq:Besselint}, for sample points in the Euclidean region. We find that the convergence is best for values close to the expansion point at $x=0$, but still reasonably good even for large values.  For complex values $z$ we do not have reference points to compare with.
However, since $P_{10}^{9}(z;0)$ and $P_{10}^{10}(z;0)$ converge in the cut plane, it makes sense to compare them against each other. In Fig.~\ref{fig:20_L_banana}, we show the logarithm of the relative error as defined in eq. \eqref{prec}, which corresponds to the number of digits that can be obtained. We find that the convergence is in general pretty good, with the worst convergence close to the branch cut on the negative real axis.

In conclusion, we have shown that thanks to the Stieltjes property, Padé approximants allow for a precise and efficient evaluation of the multi-loop `banana' integrals, even at very high orders and without relying on knowledge of their analytic form. 

\section{Discussion and outlook}\label{sec:conclusion}
In this work, we introduced a novel method for numerically bootstrapping Feynman integrals by leveraging complete monotonicity and Stieltjes properties.
We first introduced the general setup and illustrated the method using the example of the massive bubble integral. 
We also demonstrated both the practicality and robustness of the method by applying it to higher-loop cases, as a proof of principle.
This demonstrates that our method is effective for bootstrapping Feynman integrals without the need for explicit boundary conditions. The advantage of our method compared to other approaches is that it relies solely on the differential equation of an integral basis that is simple to choose. In particular, the starting differential equation matrices have rational entries.

We provided a proof that scalar Feynman integrals as a function of a single variable, while keeping all other kinematic invariants and internal masses fixed to values in the Euclidean region, for a range of parameters, are not only completely monotone, but in fact Stieltjes functions, which come with additional nice properties.
Our proof is valid for $0 < \sum\limits_{i} \nu_i - L D /2 \le 1$.
For a given integral family of interest, there typically is a choice that is within this range. The Stieltjes property implies powerful features of Padé approximants, as we reviewed. This means that the latter are well suited for analytic continuation, and there are useful convergence theorems and error bounds. 
We note that in practice it may be sufficient to have one integral (of an integral family) that is Stieltjes, in the following sense. One can then benefit from all the above properties, and use derivatives (e.g. as applied to the Padé formula) to describe the other (basis) integral in the integral family. In this way one obtains good approximations for those integrals as well.

In this paper, we focused on the case of finite Feynman integrals in integer dimensions. However, both for complete monotonicity and Stieltjes, fractional values of the dimension are allowed, so that one can also use our method in the context of dimensionally regulated integrals.
A promising future direction is to leverage these ideas to constrain the Laurent expansion in the dimensional regulator $\epsilon$.

A related approach, followed in ref.~\cite{Zeng:2023jek}, is to reconstruct the terms in the Laurent series based on numerically sampling the results for specific values of $\epsilon$.
The Stieltjes property opens up another possibility, which is to first obtain a Padé approximation, for general $\epsilon$. We have found promising evidence that the latter can be expanded in $\epsilon$ to obtain approximations of the coefficients in the Laurent expansion. We  
leave this fascinating direction for future work.

Although this work focused on Feynman integrals, we expect our method to have broader applications, including in cosmology and gravitational physics. 
This expectation comes from the fact that, as shown in ref.~\cite{Henn:2024qwe}, several objects appearing in these areas--such as cosmological correlators and Euler–Mellin integrals--satisfy the CM property. 

We find it exciting to exploit further the connection between Feynman integrals and Stieltjes functions. 
The Stieltjes property suggests a basis choice that is well adapted for numerical evaluation. 
This can be of use also in situation where differential equations are not available, thereby removing this prerequisite from our analysis.
Indeed, we have demonstrated in section \ref{sec:beyondiff} how this approach yields reliable, high-precision representations for 20-loop banana-type Feynman integrals, for example.
These ideas can be explored further along the following directions:
\begin{itemize}
\item {\it Harnessing the full power of the multivariate case.}
Feynman integrals are multivariate functions, depending simultaneously on several kinematic invariants and internal masses. In this work we analyzed them by varying one variable at a time, while keeping all remaining variables fixed in the Euclidean region. Within this framework, we proved that the resulting univariate functions are Stieltjes. This perspective allowed us to exploit the well-developed theory of univariate Stieltjes functions while still addressing the physically relevant multiscale setting. Our Padé-based approach is particularly well suited for such problems, as it provides accurate approximations without requiring the complicated special functions that typically arise in analytic representations of multi-scale Feynman integrals.

It would be interesting to explore whether genuine multivariate approaches to Stieltjes transforms and Padé approximants offer further advantages.
Several frameworks have been proposed in the approximation theory literature, see e.g. the overview
\cite{Cuyt:1983}. 
Exploring whether Feynman integrals satisfy such stronger multivariate structures is an interesting direction. 
In a related context, multi-variable extensions of complete monotonicity have been discussed in previous work by some of the present authors in ref.~\cite{Henn:2024qwe}, and further developed in ref.~\cite{Mazzucchelli:2025gyg}. 

\item {\it{Stieltjes as a basis for stable numerical evaluation.}} As we mentioned in section \ref{sec:subsecPadeexample}, employing Stieltjes functions and their derivatives as a basis of the special functions appearing in calculations can lead to important advantages over choices typically made in the literature. By definition, these functions are free from spurious divergences, and are well described by Padé approximants. This is due to a subtle interplay between properties of the transcendental functions, and their prefactors. Using such a basis could therefore be used to obtain faster and more reliable algorithms for evaluating the relevant special functions.
\item {\it{Numerical Padé interpolation.}} There are various situations relevant to collider physics where Feynman integrals are not (yet) accessible analytically, but where numerical integration methods can produce results at certain phase-space points. This typically comes with high computational cost and long runtimes. In such a situation, the Stieltjes property justifies using a multi-point Padé interpolation see refs.~\cite{Gilewicz1,Gilewicz2}. 
Our method could also be combined with other lattice reconstruction methods, such as those applied recently in ref.~\cite{Barrera:2025uin}.
\end{itemize}
In conclusion, the Stieltjes property guarantees that Padé approximants of the corresponding Feynman integrals exhibit well-established convergence, enabling efficient numerical evaluation across the cut complex plane.

 \section*{Acknowledgments}
 We are grateful to Adolfo Hilario Garcia, Jungwon Lim, and Shun-Qing Zhang for assistance with computer implementations. We also thank Leonardo de la Cruz, Maximilian Haensch, Gregory Korchemsky, Rainer Sinn, Simon Telen, Pierre Vanhove, Cristian Vergu, Qinglin Yang, and Alexander Zhiboedov 
 for interesting discussions, and to Qinglin Yang and Simone Zoia for their valuable feedback on a draft of this manuscript. 
 A preliminary version of this work was presented 
 at the SIAM Conference on Applied Algebraic Geometry (AG25), and at seminar talks at CERN and at the Max Planck Institute for Mathematics in the Sciences. We are grateful to the participants of these seminars for feedback on this work. 
 Funded by the European Union (ERC, UNIVERSE PLUS, 101118787). S.D. was supported by the ERC
Starting Grant 949279 HighPHun. Views and opinions expressed are however those of the authors only and do not necessarily reflect those of the European Union or the European Research Council Executive Agency. Neither the European Union nor the granting authority can be held responsible for them.

\appendix

\section{Feynman parametrization and the Euclidean region}
\label{sec:Feynman}

The Feynman parametrization for a scalar, $L$-loop diagram in $D$ dimensions, with propagator powers $\nu_i$, reads as follows,
\begin{align}\label{eq:FeynmanreprAppendix}
 I=\frac{\Gamma\left(\sum_i \nu_i -{L~D}/{2}\right)}{\prod_{i=1}^{L+1} \Gamma( \nu_i) } \int_0^{\infty}  \frac{\prod_i  d\alpha_i~\alpha_i^{\nu_i-1}}{\rm{GL}(1)} {U^{\sum_i\nu_i -(L+1)D/2 }}{F^{-\sum_i \nu_i +{L D}/{2} }} \,,
\end{align}
where $\alpha_i$ are the Feynman parameters.
Here the graph polynomials $U, F$ (the first and second Symanzik polynomials, respectively), are given by
\begin{align}\label{feynpara}
U &= \sum_{T \in T_1} \prod_{e_i \notin T} \alpha_i \,, \\
F &= \sum_{(T,R)\in T_2} \left( \prod_{e_i \notin(T,R)} \alpha_i\right) (-s_{T,R}) + U(\alpha) \sum_{i=1}^n \alpha_i m_i^2  \,, \label{eq:feynpara2}
\end{align}
with the kinematic variables collectively denoted by $\{x_i\}= \{-s_{T,R},m_i^2 \}$. $T_1$ in the above formula refers to a spanning tree, which is obtained from an $L$-loop graph by removing $L$ edges such that the result is a connected tree graph. $T_2$ is the two-spanning forest, obtained by removing $L+1$ edges, leading to two disconnected trees that are denoted by $T,R$.

The Euclidean region is defined as 
\begin{equation}
\mathcal{E}= \{\{x_i\}~|~ F(\{\alpha_i\};\{x_i\}) \ge 0~~ \forall~~\alpha_i \ge 0 \}  \,.  
\end{equation}
For example, for the massive bubble integral considered in the main text, we have 
\begin{equation}
\begin{aligned}\label{eq:bubba1}
 F(\alpha_1,\alpha_2;s,m_1^2,m_2^2) =& \alpha_1\alpha_2 (-s)+ (\alpha_1+\alpha_2)(\alpha_1 m_1^2+\alpha_2 m_2^2)  \\
  \ge& \alpha_1\alpha_2((m_1+m_2)^2-s)\,.
 \end{aligned}
\end{equation}
with the inequality being saturated for $\alpha_1=m_2,  \alpha_2=m_1$. We find 
\begin{align}\label{eq:bubbleLandau}
 \mathcal{E}(s,m_1^2,m_2^2)&= \{(s,m_1,m_2)~|~ s \le (m_1+m_2)^2 \}  \,.
\end{align} 
We note that for $x=-s$, and setting $m_1=m_2=1$,  (\ref{eq:bubbleLandau}) gives $x>-4$. 

Another example is the massive box integral with equal internal masses. Here we have 
\begin{equation}
\begin{aligned}\label{eq:massbox}
F(\alpha_1,...,\alpha_4;s,t,m^2) =& \alpha_1 \alpha_3 (-s)+   \alpha_2 \alpha_4 (-t)+(\alpha_1+\alpha_2+\alpha_3+\alpha_4)^2~m^2 \\
 \ge& \alpha_1~\alpha_3(4m^2-s)+\alpha_2~\alpha_4(4m^2-t),
\end{aligned}
\end{equation}
with the inequality being saturated for $\alpha_1, \alpha_3=0,\alpha_2=\alpha_4$ or $\alpha_2, \alpha_4=0,\alpha_1=\alpha_3$. We find 
\begin{eqnarray}\label{eqn:massbocLandau}
\mathcal{E}(s,t,m^2)&=&\{(s,t,m)~|~s\leq 4m^2,~t \leq 4m^2\} \,.  
\end{eqnarray}
The Euclidean region can be geometrically characterized as a {copositive cone}, see ref.~\cite{Sturmfels:2025wpg}.

\begin{figure}[t]
    \centering
    \subfloat[\label{fig:appb}]{\includegraphics[width=0.4\textwidth]{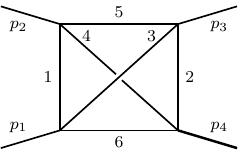}}
    \hspace{4em}
    \subfloat[\label{fig:appa}]{\includegraphics[width=0.4\textwidth]{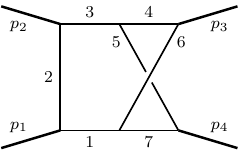}}\\
    \caption{Examples of non-planar Feynman integrals for which the Symanzik polynomial can be put in the form of eq. (\ref{eq:technicalcondition}). (a) Three-loop $K_{4}$ graph.
    (b) Non-planar double box. }
    \label{fig:NPLexamples}
\end{figure} 

Let us now see how the second Symanzik polynomial can be expressed in the form of eq. %\eqref{eq:Symanzik2} or 
\eqref{eq:technicalcondition}, which was an assumption in our proof of the Stieltjes property.
%in section \ref{sec:FeynStielt}. 

For the massive bubble, the $F(\{\alpha_i\})$  in eq.~\eqref{eq:bubba1} can be written as 
\begin{align}\label{eq:manifestpositivity}
    F(\{\alpha_i\})=\sum_i x_i A_i\,,
\end{align}
with manifestly positive coefficients $A_i \ge 0$, 
and a set of {\it unconstrained} kinematic variables $x_i$.\footnote{Note that here we denote by $x_i$ a set of independent variables, which is different from the in-line definition after eq. (\ref{eq:feynpara2}), which in turn refers to the (dependent) variables in the graph-theoretical definition. We hope that this does not lead to confusion.}
This can be achieved by choosing
\begin{equation}
    \{x_i \}=\{-s, m_1^2, m_2^2\} \quad \text{or}\quad \{x_i \}= \{m_1^2+m_2^2-s, m_1^2, m_2^2\} \,.
\end{equation}
In both cases the positivity of the $A_i$ establishes the Stieltjes property, since the requirement in eq.~\eqref{eq:manifestpositivity} is satisfied. For the former choice we obtain 
\begin{equation}
    A_1=\alpha_1\alpha_2,\,\quad  A_2=\alpha_1(\alpha_1+\alpha_2),\, \quad  A_3=\alpha_2(\alpha_1+\alpha_2)\,,
\end{equation}
 while for the latter choice we find 
 \begin{equation}
     A_1=\alpha_1\alpha_2,\, \quad   A_2=\alpha_1^2\, \quad   A_3=\alpha_2^2 \,.
 \end{equation}
Finally, let us mention the choice of variables $\{x_i\}= \{(m_1+m_2)^2-s, m_1, m_2\}$. In this case, singeling out $x=(m_1+m_2)^2-s$, we obtain $F(\{\alpha_i\}) = A\,x + B$ with $A=\alpha_1\alpha_2$ and a non-linear term $B=(\alpha_1 m_1-\alpha_2 m_2)^2$, which is of the form given in eq.~\eqref{eq:technicalcondition} and thus establishes the Stieltjes property in the variable $x=(m_1+m_2)^2-s$.

We now give examples to illustrate
that obtaining the form in eq.~\eqref{eq:manifestpositivity} is not restricted to planar integrals.
For the three-loop Feynman integral corresponding to the $K_4$ graph with one off-shell leg, shown in Fig.~\ref{fig:NPLexamples}(a), one can obtain the manifestly positive linear form of the second Symanzik polynomial by making the choice of variables 
\begin{equation}
    \{ x_{i} \}=\{-s,\,-t,\,-u\} \,,
\end{equation}
for which one finds \cite{Henn:2013nsa}
\begin{equation}
\begin{aligned}
    A_1=&\alpha_2\alpha_3 (\alpha_1\alpha_4 + \alpha_1\alpha_6 + \alpha_4 \alpha_6 + \alpha_5 \alpha_6) \,, \\
    A_2=& \alpha_3\alpha_6 (\alpha_1\alpha_2+ \alpha_2\alpha_4 + \alpha_2\alpha_5 + \alpha_4 \alpha_5) \,, \\
    A_3=& \alpha_2\alpha_6(\alpha_1\alpha_3 + \alpha_3 \alpha_4 + \alpha_1\alpha_5 + \alpha_3 \alpha_5) \,. 
\end{aligned}
\end{equation}
For the non-planar double box, shown in Fig.~\ref{fig:NPLexamples}(b), the second Symanzik polynomial can be put in the form of eq. \eqref{eq:manifestpositivity} by choosing independent variables as follows,
\begin{align}
    \{ x_i \}=\{-s+p_1^2+p_2^2,\,-t+p_2^2+p_3^2,\,-u+p_1^2+p_3^2,\,
    -p_1^2,\,
    -p_2^2,\,
    -p_3^2\} \,,
\end{align}
which gives the manifestly positive coefficients
\begin{equation}
\begin{aligned}
    A_1\,=&\alpha_1\alpha_5 \alpha_6 + \alpha_2 \alpha_5 \alpha_6 + \alpha_3 \alpha_4 \alpha_7 + \alpha_1\alpha_5 \alpha_7 + \alpha_2 \alpha_5 \alpha_7 + \alpha_3 \alpha_5 \alpha_7 + 
    \alpha_4 \alpha_5 \alpha_7 \nonumber \\ 
    &+ \alpha_5 \alpha_6 \alpha_7  \,, \\
    A_2\,=&\alpha_1\alpha_3 \alpha_4 + \alpha_1\alpha_3 \alpha_5 + \alpha_1\alpha_4 \alpha_5 + 
    \alpha_1\alpha_3 \alpha_6 + \alpha_1\alpha_5 \alpha_6 + \alpha_1\alpha_3 \alpha_7 + \alpha_3 \alpha_4 \alpha_7 \,, \nonumber  \\ 
    &+ \alpha_1\alpha_5 \alpha_7+ \alpha_2 \alpha_5 \alpha_7 + 
    \alpha_3 \alpha_5 \alpha_7 + \alpha_4 \alpha_5 \alpha_7 + \alpha_3 \alpha_6 \alpha_7 + \alpha_5 \alpha_6 \alpha_7 \,, \\
    A_3\,=&\alpha_1\alpha_5 \alpha_6 + 
    \alpha_2 \alpha_4 \alpha_7 + \alpha_3 \alpha_4 \alpha_7+ \alpha_1\alpha_5 \alpha_7 + \alpha_2 \alpha_5 \alpha_7 + \alpha_3 \alpha_5 \alpha_7 + \alpha_4 \alpha_5 \alpha_7 \nonumber \\ 
    &+ 
    \alpha_5 \alpha_6 \alpha_7  \,,\\
    A_4\,=&\alpha_1\alpha_2 \alpha_4 + \alpha_1\alpha_2 \alpha_5 + \alpha_1\alpha_2 \alpha_6 + \alpha_1\alpha_5 \alpha_6 + 
    \alpha_2 \alpha_5 \alpha_6 + \alpha_1\alpha_2 \alpha_7 + \alpha_2 \alpha_4 \alpha_7 \nonumber \\
    &+ \alpha_3 \alpha_4 \alpha_7 + \alpha_1\alpha_5 \alpha_7 + \alpha_2 \alpha_5 \alpha_7 + 
    \alpha_3 \alpha_5 \alpha_7 + \alpha_4 \alpha_5 \alpha_7 + \alpha_2 \alpha_6 \alpha_7 + \alpha_5 \alpha_6 \alpha_7  \,,\\
    A_5\,=&\alpha_1\alpha_3 \alpha_4 + 
    \alpha_2 \alpha_3 \alpha_4 + \alpha_1\alpha_3 \alpha_5 + \alpha_2 \alpha_3 \alpha_5 + \alpha_1\alpha_4 \alpha_5 + \alpha_2 \alpha_4 \alpha_5 + \alpha_1\alpha_3 \alpha_6 \nonumber \\
    &+ 
    \alpha_2 \alpha_3 \alpha_6 + \alpha_1\alpha_5 \alpha_6 + \alpha_2 \alpha_5 \alpha_6 + \alpha_1\alpha_3 \alpha_7 + \alpha_2 \alpha_3 \alpha_7 + \alpha_3 \alpha_4 \alpha_7 + 
    \alpha_1\alpha_5 \alpha_7  \nonumber\\
    & + \alpha_2 \alpha_5 \alpha_7 + \alpha_3 \alpha_5 \alpha_7 + \alpha_4 \alpha_5 \alpha_7 + \alpha_3 \alpha_6 \alpha_7 + 
    \alpha_5 \alpha_6 \alpha_7  \,,\\
    A_6\,=&\alpha_1\alpha_3 \alpha_4 + \alpha_1\alpha_3 \alpha_5 + \alpha_1\alpha_4 \alpha_5 + \alpha_1\alpha_3 \alpha_6 + 
    \alpha_1\alpha_4 \alpha_6 + \alpha_2 \alpha_4 \alpha_6 + \alpha_3 \alpha_4 \alpha_6 \nonumber \\
    &+ \alpha_1\alpha_5 \alpha_6+ \alpha_3 \alpha_5 \alpha_6 + \alpha_4 \alpha_5 \alpha_6 + 
    \alpha_1\alpha_3 \alpha_7 + \alpha_1\alpha_4 \alpha_7 + \alpha_2 \alpha_4 \alpha_7 + \alpha_3 \alpha_4 \alpha_7 \nonumber \\
    &+ \alpha_1\alpha_5 \alpha_7 + \alpha_2 \alpha_5 \alpha_7 + 
    \alpha_3 \alpha_5 \alpha_7 + \alpha_4 \alpha_5 \alpha_7 + \alpha_3 \alpha_6 \alpha_7 + \alpha_4 \alpha_6 \alpha_7 + \alpha_5 \alpha_6 \alpha_7 \,.
\end{aligned}
\end{equation}
It is an interesting question how to find such variables choices systematically.

\bibliographystyle{JHEP} 
\bibliography{ff.bib}

\end{document}